\documentclass[prx,aps,twocolumn,floatfix,superscriptaddress,nofootinbib]{revtex4-2}
\usepackage{booktabs}
\usepackage{graphicx}
\usepackage{braket}
\usepackage{stix}
\usepackage{xcolor}

\usepackage{physics}

\definecolor{reviewblue}{RGB}{0,85,160}
\definecolor{reviewgreen}{RGB}{0,150,0}

\usepackage[hidelinks]{hyperref}
\usepackage[nameinlink,noabbrev]{cleveref}
\usepackage{tasks}
\usepackage{adjustbox} 

\makeatletter
\usepackage{dcolumn}
\usepackage{soul}
\usepackage{bm}
\usepackage{hyperref}
\usepackage{amsthm}
\usepackage{enumitem}
\usepackage{amsmath}

\usepackage{xr-hyper}

\usepackage[caption=false]{subfig}
\hypersetup{colorlinks,
	citecolor=blue
}

\usepackage{mathtools}
\usepackage{apptools}
\usepackage{nameref}
\AtAppendix{\counterwithin{lemma}{section}}

\makeatletter
\newcommand*{\addFileDependency}[1]{
  \typeout{(#1)}
  \@addtofilelist{#1}
  \IfFileExists{#1}{}{\typeout{No file #1.}}
}
\makeatother

\crefname{section}{Appendix}{Appendices}
\Crefname{section}{Appendix}{Appendices}

\renewcommand\thesection{\Roman{section}}

\begin{document}

\hyphenation{Dzyaloshinskii-Moriya}
\hyphenation{chirality-renormalized}
\hyphenation{Dzyaloshinskii}
\hyphenation{inter-chain}
\hyphenation{di-mer-ized}

\preprint{APS/123-QED}

\title{\textbf{Topological Edge States from Molecular Chirality: A General Framework for Dimerized Dipolar Arrays} 
}%

\author{Muhammad Arsalan Ali Akbar}
\affiliation{Department of Electrical and Computer Engineering, North Carolina State University, Raleigh, NC 27606}
\affiliation{ Department of Chemistry, North Carolina State University, Raleigh, North Carolina 27695}

\author{Mohsin Raza}
\affiliation{Department of Electrical and Computer Engineering, North Carolina State University, Raleigh, NC 27606 }

\author{Sabre Kais}
\email{skais@ncsu.edu}
\affiliation{Department of Electrical and Computer Engineering, North Carolina State University, Raleigh, NC 27606}
\affiliation{ Department of Chemistry, North Carolina State University, Raleigh, North Carolina 27695}

\begin{abstract}
We establish a general theoretical framework for realizing topological
edge states in dimerized arrays of chiral dipolar molecules and
demonstrate that molecular handedness provides a natural and tunable
route to SSH-like topology in an interacting one-dimensional setting.
Starting from an effective spin-$\tfrac{1}{2}$ model generated by
Stark-dressed chiral molecules, we introduce bond dimerization and
show that the chirality-induced Dzyaloshinskii--Moriya interaction
amplifies the effective hopping amplitudes and enlarges the bulk
topological gap relative to an achiral chain of equivalent dipole
strength. Using self-consistent mean-field theory with periodic- and
open-boundary calculations, we map out the trivial, critical, and
topological regimes through bulk spectra, complex-plane winding, and
boundary-localized probability densities. A central result is that
the two in-gap boundary modes carry \emph{opposite molecular
chirality}: the left edge state localizes on a left-handed molecule
and the right edge state on a right-handed molecule, a
stereochemical labeling with no analogue in conventional SSH
implementations. The two-leg ladder extension supports a richer
four-band bulk structure and a rung-split edge sector whose
robustness is characterized by a continuous sweep of the interchain
coupling. All results are expressed in dimensionless units of the
reference hopping scale $t_0$, making the framework directly
applicable to any dipolar molecular platform --- from bialkali polar
molecules at MHz coupling scales to future arrays of ultracold
chiral polyatomic species. These findings establish dimerized chiral
molecular arrays as a controllable and chirality-addressable
platform for quasi-one-dimensional topological quantum matter.
\end{abstract}

\maketitle
\section{Introduction}
\label{sec:introduction}

Topological phases of matter have reshaped the modern understanding of
quantum systems by showing that phases can be distinguished not only by
local order parameters, but also by global geometric and topological
properties of their quantum states~\cite{HasanKane2010,QiZhang2011,
Chiu2016,Bansil2016}. In one dimension, this idea is realized in a
particularly transparent form by the Su--Schrieffer--Heeger (SSH)
model, where alternating nearest-neighbor bonds generate two insulating
phases separated by a bulk-gap closing point~\cite{SSH1979,
Heeger1988,Asboth2016}. The nontrivial phase supports
boundary-localized modes whose existence is fixed by the winding of the
bulk hopping function, making the SSH chain one of the simplest
realizations of the bulk--boundary correspondence. Owing to this
minimal structure, SSH physics has become a central reference point for
topological phenomena in condensed matter, cold-atom systems, photonic
lattices, and other synthetic platforms~\cite{Atala2013,Cooper2019,
Ozawa2019,Lu2014,Meier2016}.

\setlength{\fboxrule}{0.6pt}
\setlength{\fboxsep}{6pt}

\begin{figure*}[!t]
\centering
\fbox{%
  \begin{minipage}{\dimexpr\textwidth-2\fboxsep-2\fboxrule\relax}
  \centering
  \vspace{2pt}

  \begin{minipage}[t]{0.48\textwidth}
      \centering
      \includegraphics[width=\linewidth]{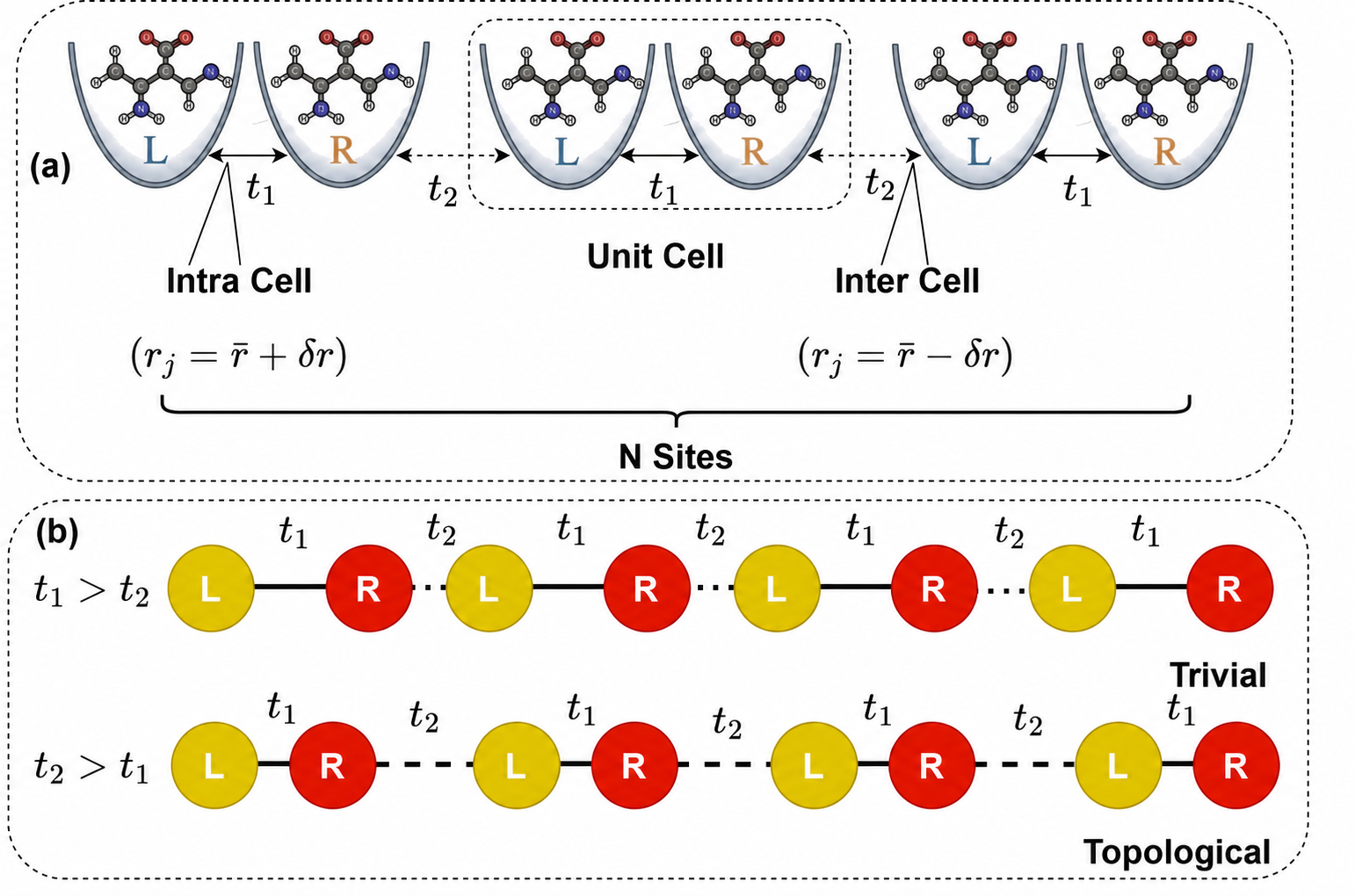}
  \end{minipage}
  \hfill
  \begin{minipage}[t]{0.48\textwidth}
      \centering
      \includegraphics[width=\linewidth]{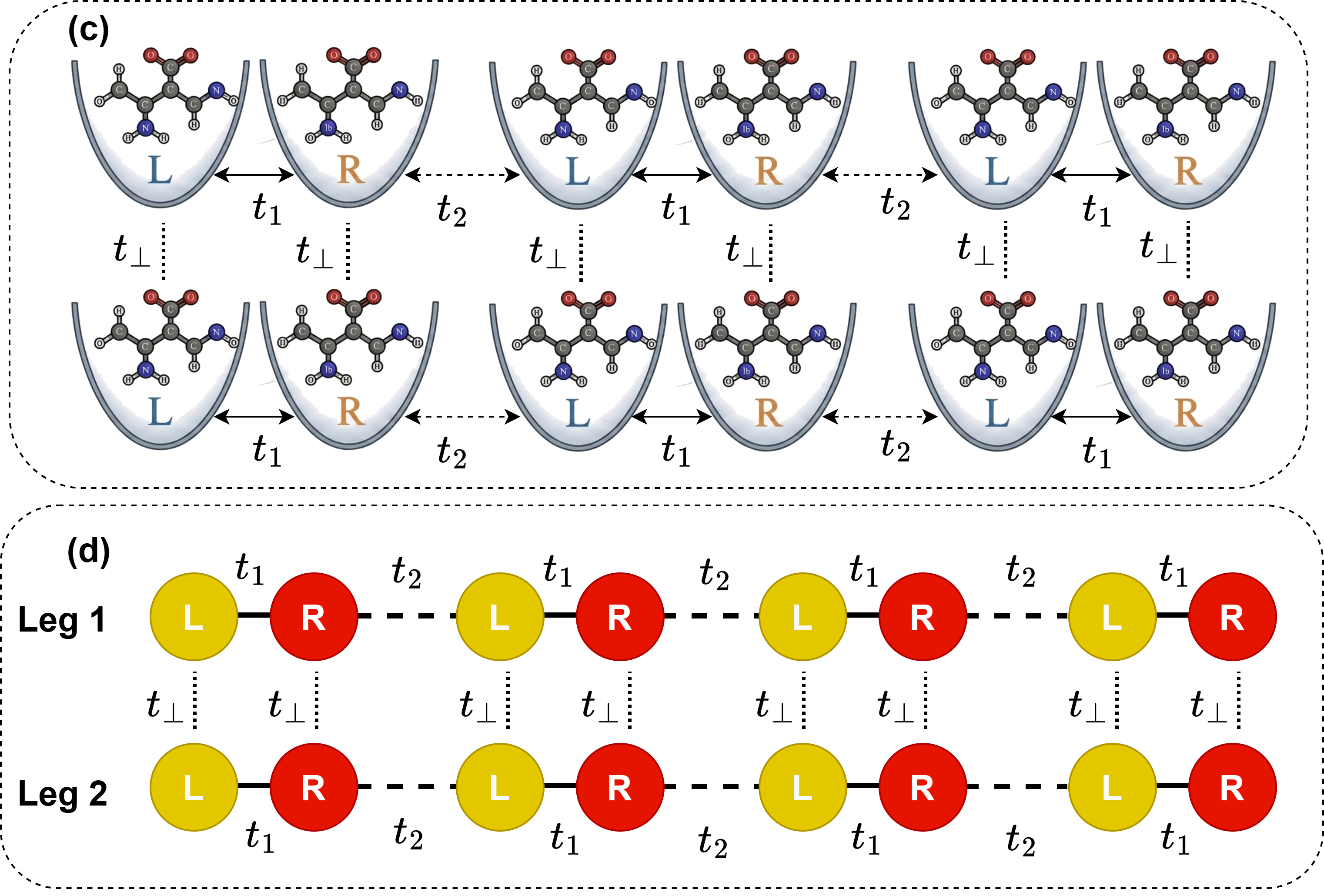}
  \end{minipage}

  \vspace{2pt}
  \end{minipage}%
}

\caption{Physical platform and effective models studied in this work.
(a)~Stark-dressed chiral molecular chain realized in an optical tweezer
array. Alternating left-handed (L, blue) and right-handed (R, orange)
enantiomers are individually trapped in optical potential wells. With
the dimerization convention used in the text,
$r_j=\bar r-(-1)^j\delta r$, odd bonds are intracell bonds and even
bonds are intercell bonds. Thus the intracell L--R bond carries hopping
$t_1$ and has separation $r_j=\bar r+\delta r$, while the intercell bond
connecting the R site of one unit cell to the L site of the next carries
hopping $t_2$ and has separation $r_j=\bar r-\delta r$. For positive
dimerization, the intercell bond is shorter and therefore stronger,
giving $t_2>t_1$. The dashed rectangle marks one unit cell.
(b)~Effective dimerized SSH chain obtained after the gauge rotation and
Jordan--Wigner mapping of panel~(a). Yellow (L) and red (R) filled
circles denote the two sublattice sites. In the trivial termination
($t_1>t_2$), the strong bond lies inside the unit cell and the winding is
$\nu=0$. In the topological termination ($t_2>t_1$), the strong bond
lies between neighboring unit cells and the winding is $|\nu|=1$; the
outermost L site at the left boundary and R site at the right boundary
are left unpaired, producing the two molecular-chirality-labeled edge
states. (c)~Two-leg ladder extension of the chiral molecular array,
formed by placing a second identical L--R chain at transverse separation
$r_\perp$. Corresponding sites on the two legs are coupled by an
interchain hopping $t_\perp\propto r_\perp^{-3}$. (d)~Effective two-leg
ladder model. Each leg carries the same alternating $t_1$--$t_2$
intrachain pattern, while vertical rung hoppings $t_\perp$ hybridize the
two chains. In the topological regime, the ladder supports four in-gap
boundary modes, two at each end, which are split by the rung
hybridization.}
\label{fig:overview_model}
\end{figure*}

Molecular quantum platforms provide a complementary route to engineered
many-body Hamiltonians with microscopic tunability. Ultracold polar
molecules possess long-lived internal states, strong electric
dipole--dipole interactions, and external-field control, making them
promising systems for quantum simulation~\cite{Carr2009,Bohn2017,kais-2024,sabre-2011,kais-2024,kais-2013}, quantum information processing~\cite{Micheli2006,Sawant2020}, and
precision measurements~\cite{Anderegg2023}. Recent progress in optical
tweezer arrays has further strengthened this direction by enabling
site-resolved preparation, rearrangement, and coherent control of
individual molecules~\cite{Kaufman2021,Anderegg2019,Ruttley2024}. In
several recent experiments, such platforms have also enabled
deterministic entanglement of molecular qubits~\cite{Holland2023,
Bao2023,Ruttley2025}. These developments suggest that molecular arrays
can serve as highly controllable platforms for realizing lattice
Hamiltonians whose couplings are set by internal molecular structure and
programmable geometry.

Within this broader setting, chiral molecules are especially appealing
because molecular handedness can enter directly into the effective
interaction structure. In a static electric field, rotational levels are
mixed by the Stark effect, producing dressed states that can be selected
as an effective pseudo-spin-$1/2$ subspace. As recently shown for
Stark-dressed chiral molecular arrays, projection of the dipole--dipole
interaction onto this subspace generates an effective spin Hamiltonian
with anisotropic exchange and a chirality-induced Dzyaloshinskii--Moriya
(DM) interaction~\cite{Akbar2024,Dzyaloshinsky1958,Moriya1960}. This
provides a microscopic connection between stereochemical structure and
interacting quantum magnetism. However, whether such chiral molecular
arrays can also host topological boundary states has remained largely
unexplored.

A natural route to this question is to impose bond dimerization on the
nearest-neighbor molecular spacing. Since the dipolar interaction scale
varies as $r^{-3}$, an alternating intermolecular distance directly
produces alternating exchange amplitudes. A local gauge transformation
removes the DM phase from the open-chain bulk and absorbs its effect
into the chirality-renormalized transverse exchange $\left(\widetilde J_{xy,j}=\sqrt{J_{xy,j}^{2}+D_j^{2}}\right)$.
The resulting gauge-rotated model maps, through the Jordan--Wigner
transformation, to an interacting SSH-like fermionic chain with
chirality-renormalized hopping amplitudes. In this construction, the
topological distinction is controlled by bond dimerization, while
molecular chirality fixes the microscopic hopping scale and gives a
stereochemical identity to the terminal sites.

The physical construction studied in this work is summarized in
Fig.~\ref{fig:overview_model}. Fig.~\ref{fig:overview_model}(a)
shows a Stark-dressed chiral molecular chain with alternating
left-handed (L) and right-handed (R) molecules.
Fig.~\ref{fig:overview_model}(b) shows the corresponding effective
dimerized SSH chain obtained after the gauge rotation and
Jordan--Wigner mapping. In the convention used throughout this work,
odd bonds are intracell bonds with hopping $t_1$, even bonds are
intercell bonds with hopping $t_2$, and positive dimerization gives
$t_2>t_1$, realizing the topological termination.
Fig.~\ref{fig:overview_model}(c) extends the construction to two
coupled dimerized chains, and Fig.~\ref{fig:overview_model}(d) shows
the corresponding effective two-leg ladder. This ladder geometry provides the simplest
quasi-one-dimensional extension of the single-chain problem and allows
us to examine how interchain hybridization modifies both the bulk bands
and the boundary-state sector.

In this work, we develop a unified theoretical framework for dimerized
chiral dipolar arrays in both single-chain and two-leg ladder
geometries. Starting from the effective chiral molecular spin model, we
derive the interacting SSH-type fermionic Hamiltonian, perform a
self-consistent Hartree--Fock reduction at half filling, and analyze the
resulting bulk and boundary properties. The bulk phases are diagnosed
using periodic-boundary spectra and the winding of the complex hopping
function, while the boundary physics is examined through open-boundary
spectra and site-resolved probability densities. This combination of
diagnostics allows us to map the trivial, critical, and topological
regimes within a single framework.

The central physical result is that the topological phase supports
boundary-localized modes whose dominant molecular character is tied to
the handedness of the terminal molecules. For the topological
termination, the left edge mode is localized predominantly on an
L-molecule site, while the right edge mode is localized predominantly on
an R-molecule site. These modes are therefore ordinary SSH edge states
in their topological origin, but chirality-addressable in their
molecular realization. This stereochemical labeling has no direct
analogue in conventional achiral SSH chains, where boundary sites carry
no molecular handedness. Thus, the edge states are not only
topologically selected by the dimerized hopping pattern, but also
distinguishable through the molecular handedness of the sites on which
they localize. The two-leg ladder further enriches the boundary sector:
two topological chains contribute four in-gap edge modes, and the rung
coupling splits them into bonding and antibonding pairs while preserving
their boundary-localized character within a well-defined protection
window.

The broader significance of this construction is that it connects two
ideas that are usually treated separately. Dimerized one-dimensional
lattices provide a minimal route to topological boundary modes, whereas
chiral molecular arrays provide a microscopic platform in which
stereochemistry enters the effective exchange structure. Combining these
ingredients yields a controllable SSH-like molecular system in which
topology, dipolar interactions, and molecular handedness are interwoven.

The paper is organized as follows. In Sec.~\ref{sec:model}, we introduce
the effective chiral molecular Hamiltonian and derive its dimerized
single-chain and ladder descriptions. In Sec.~\ref{sec:results}, we
present the bulk spectra, winding structure, open-boundary spectra,
real-space edge-state profiles, and rung-coupling robustness analysis.
Finally, in Sec.~\ref{sec:conclusion}, we summarize the main results and
discuss possible extensions beyond the present mean-field treatment.

\section{Model and Method}
\label{sec:model}

We consider a dimerized array of Stark-dressed chiral molecules described at low energies by an effective pseudo-spin-$1/2$ model~\cite{Akbar2024}. The pseudo-spin basis is formed by two selected dressed rotational states of each molecule, and projection of the dipole--dipole interaction onto this two-level subspace gives a chiral XXZ spin chain. The resulting spin Hamiltonian contains symmetric transverse exchange, a Dzyaloshinskii--Moriya (DM) interaction, a longitudinal Ising interaction, and an effective field. In the present work, we introduce bond dimerization into this chiral molecular chain and use the resulting model to construct an interacting SSH-type fermionic system and its two-leg ladder extension.

\subsection{From the chiral spin chain to an interacting SSH model}

In the Pauli-matrix convention, the effective open-chain spin Hamiltonian is~\cite{Akbar2024}
\begin{align}
\hat H_{\mathrm{spin}}
&=
\sum_{j=1}^{N-1}
\Big[
J_{xy,j}
(\hat\sigma_j^x\hat\sigma_{j+1}^x+
 \hat\sigma_j^y\hat\sigma_{j+1}^y)
\nonumber\\
&\quad
-
D_j
(\hat\sigma_j^x\hat\sigma_{j+1}^y-
 \hat\sigma_j^y\hat\sigma_{j+1}^x)
+
J_{z,j}\hat\sigma_j^z\hat\sigma_{j+1}^z
\Big]
+
h\sum_{j=1}^{N}\hat\sigma_j^z .
\label{eq:model_spin_hamiltonian}
\end{align}
Here \(\hat\sigma_j^\alpha\) are Pauli matrices acting on site \(j\). The first term is the symmetric transverse exchange, the second is the chirality-induced DM interaction, the third is the Ising anisotropy, and the final term is the effective longitudinal field. We evaluate the field at the mean separation, \(h\equiv h(\bar r)\), while the exchange couplings retain their bond dependence through \(r_j\).

The microscopic couplings are
\begin{align}
J_{xy,j}
&=
-\frac{\Omega(r_j)}{2}\Re(C_{d_1}),
\\
D_j
&=
+\frac{\Omega(r_j)}{2}\Im(C_{d_1}),
\\
J_{z,j}
&=
\frac{\Omega(r_j)}{4}
\big[(C_2+C_3)-(C_1+C_4)\big],
\\
h(\bar r)
&=
\frac{
2(E_\uparrow-E_\downarrow)
+
\Omega(\bar r)(C_1-C_4)
}{4}.
\end{align}
The dipolar energy scale is
\begin{equation}
\Omega(r_j)
=
\frac{1}{h_{\mathrm P}}
\frac{|\mathbf d|^2}{4\pi\epsilon_0 r_j^3},
\qquad
\Omega(r_j)\propto r_j^{-3}.
\end{equation}
Here \(h_{\mathrm P}\) is Planck's constant, \(\mathbf d\) is the molecular dipole moment, \(\epsilon_0\) is the vacuum permittivity, \(C_{d_1}\) and \(C_i\) are dressed-state dipole matrix elements, and \(E_\uparrow,E_\downarrow\) are the dressed-state energies.

The important structural point is that \(J_{xy,j}\) and \(D_j\) contain the same distance-dependent factor \(\Omega(r_j)\). Therefore,
\[
\frac{D_j}{J_{xy,j}}
=
-
\frac{\Im(C_{d_1})}{\Re(C_{d_1})},
\]
which is independent of the bond length. This allows the DM phase to be removed from the open-chain bulk by a site-dependent rotation about the \(z\)-axis. Chirality is not discarded by this transformation; it is absorbed into the magnitude of the transverse exchange,
\[
\widetilde J_{xy,j}
=
\sqrt{J_{xy,j}^2+D_j^2}.
\]
The explicit gauge transformation and the associated boundary-condition discussion are given in Appendix~\ref{app:gauge_jw}.

After the gauge rotation, the model is mapped to spinless fermions using the Jordan--Wigner transformation. With the Pauli convention used here,
\[
\hat\sigma_j^z=2\hat n_j-1,
\qquad
\hat n_j=\hat c_j^\dagger\hat c_j .
\]
For nearest-neighbor terms on an open chain, the Jordan--Wigner strings cancel exactly. The mapping gives a hopping term, a nearest-neighbor density interaction, linear density terms, and a constant energy shift. In the bulk-effective treatment we keep the dimerization explicitly in the transverse hopping amplitudes and take the longitudinal interaction at the mean separation,
\[
J_z\equiv J_z(\bar r).
\]
The corresponding bulk-effective fermionic Hamiltonian is
\begin{align}
\hat H_{\mathrm{bulk}}
&=
\sum_{j=1}^{N-1}
t_j
(\hat c_j^\dagger\hat c_{j+1}+\mathrm{h.c.})
+
4J_z
\sum_{j=1}^{N-1}
\hat n_j\hat n_{j+1}
\nonumber\\
&\quad
+
\mu_{\mathrm{bulk}}
\sum_{j=1}^{N}
\hat n_j
+
E_{\mathrm{const}} .
\label{eq:model_bulk_fermion}
\end{align}
The parameters in Eq.~\eqref{eq:model_bulk_fermion} are
\begin{align}
t_j &= 2\widetilde{J}_{xy,j}, \qquad
\mu_{\mathrm{bulk}} = 2h - 4J_z,
\nonumber\\
E_{\mathrm{const}} &= J_z(N-1) - hN .
\label{eq:model_bulk_parameters}
\end{align}
The hopping \(t_j\) is generated by the gauge-renormalized transverse exchange, and the factor of \(2\) follows from the Pauli-matrix convention. The term \(4J_z\hat n_j\hat n_{j+1}\) is the nearest-neighbor density--density interaction inherited from the Ising coupling. The coefficient \(\mu_{\mathrm{bulk}}\) is the bulk linear density coefficient obtained after combining the longitudinal field with the linear density terms produced by the Ising interaction. The constant \(E_{\mathrm{const}}\) shifts all eigenvalues uniformly and does not affect eigenstates, gaps, winding numbers, or edge-state localization.

For a finite open chain, the exact Jordan--Wigner mapping also produces a boundary density correction,
\[
2J_z(\hat n_1+\hat n_N),
\]
because the two boundary sites belong to only one bond. This term is omitted in Eq.~\eqref{eq:model_bulk_fermion}, since the model and bulk topology are governed by the thermodynamic interior. The full open-chain accounting is given in Appendix~\ref{app:gauge_jw}.

Dimerization is introduced by alternating the intermolecular separation as
\begin{equation}
r_j=\bar r-(-1)^j\delta r .
\label{eq:model_dimerization}
\end{equation}
With this convention, odd bonds have \(r_j=\bar r+\delta r\), while even bonds have \(r_j=\bar r-\delta r\). Since \(\Omega(r)\propto r^{-3}\), positive \(\delta r\) makes the even bonds stronger than the odd bonds. The two hopping amplitudes are therefore
\begin{align}
t_1
&=
2\widetilde J_{xy}(\bar r)
\left(
\frac{\bar r}{\bar r+\delta r}
\right)^3,
\\
t_2
&=
2\widetilde J_{xy}(\bar r)
\left(
\frac{\bar r}{\bar r-\delta r}
\right)^3 .
\end{align}
We choose odd bonds as intracell bonds and even bonds as intercell bonds,
\[
t_j=
\begin{cases}
t_1, & j\ \mathrm{odd},\\
t_2, & j\ \mathrm{even}.
\end{cases}
\]
Thus, for \(\delta r>0\), the intercell hopping satisfies \(t_2>t_1\), which is the topological SSH pattern for the termination used in this work. After the gauge rotation and Jordan--Wigner mapping, the dimerized chiral molecular spin chain is therefore represented by an interacting SSH-type fermionic model.

\subsection{Hartree--Fock reduction}

The density--density interaction in Eq.~\eqref{eq:model_bulk_fermion} prevents a direct single-particle treatment. We therefore decouple the quartic term at the Hartree--Fock level at half filling. The two inequivalent SSH bonds are assigned independent bond-order parameters,
\begin{align}
\chi_1
&=
\frac{1}{\mathcal N_1}
\sum_{j\ \mathrm{odd}}
\langle
\hat c_j^\dagger\hat c_{j+1}
\rangle,
\\
\chi_2
&=
\frac{1}{\mathcal N_2}
\sum_{j\ \mathrm{even}}
\langle
\hat c_j^\dagger\hat c_{j+1}
\rangle .
\end{align}
Here \(\mathcal N_1\) and \(\mathcal N_2\) are the numbers of odd and even bonds included in the corresponding averages. The quantities \(\chi_1\) and \(\chi_2\) measure the quantum coherence across the two SSH bond types and correspond, in the spin language, to nearest-neighbor flip-flop correlations.

The Hartree channel shifts the density term, while the Fock channel renormalizes the hopping amplitudes. Combining the exact bulk linear density coefficient with the Hartree contribution gives
\[
\mu_{\mathrm{eff}}
=
\mu_{\mathrm{bulk}}+4J_z
=
2h.
\]
This term is a uniform spectral shift. It is retained when absolute energies are quoted but subtracted when computing topological quantities. The Fock channel gives
\begin{equation}
t_1^{\mathrm{eff}}
=
t_1-4J_z\chi_1,
\qquad
t_2^{\mathrm{eff}}
=
t_2-4J_z\chi_2 .
\label{eq:model_effective_hoppings}
\end{equation}
After subtracting the uniform shift and dropping the constant energy offset, the reduced mean-field Hamiltonian becomes
\begin{align}
\hat H_{\mathrm{MF}}'
&=
\sum_{j\ \mathrm{odd}}
t_1^{\mathrm{eff}}
(\hat c_j^\dagger\hat c_{j+1}+\mathrm{h.c.})
\nonumber\\
&\quad+
\sum_{j\ \mathrm{even}}
t_2^{\mathrm{eff}}
(\hat c_j^\dagger\hat c_{j+1}+\mathrm{h.c.}) .
\label{eq:model_mf_hamiltonian}
\end{align}
The bond orders \(\chi_1\) and \(\chi_2\) are computed self-consistently. Under open boundary conditions, Eq.~\eqref{eq:model_mf_hamiltonian} is diagonalized in real space, the lowest \(N/2\) single-particle states are occupied, and the bond orders are recomputed until convergence. The Hartree--Fock derivation is given in Appendix~\ref{app:mf_decoupling}.

\subsection{Bulk topology and open-boundary diagnostics}

After the self-consistency loop has converged, the interacting problem is reduced to an effective SSH chain with renormalized hoppings \(t_1^{\mathrm{eff}}\) and \(t_2^{\mathrm{eff}}\). The bulk topological properties are most naturally obtained under periodic boundary conditions, where translational symmetry allows the Hamiltonian to be written in momentum space. In the two-site unit-cell representation, the only nonzero matrix element connecting the two sublattices is the complex SSH function
\begin{equation}
q(k) = t_1^{\mathrm{eff}} + t_2^{\mathrm{eff}}\,e^{-ik}.
\label{eq:model_hk}
\end{equation}
Thus, after subtracting the uniform density shift, the Bloch Hamiltonian takes the chiral form
\begin{equation}
\mathcal H(k)
=
\begin{pmatrix}
0 & q(k)\\
q^*(k) & 0
\end{pmatrix}.
\label{eq:model_bloch_hamiltonian}
\end{equation}
If absolute energies are required, the term \(\mu_{\mathrm{eff}}\mathbb I\) is added back to Eq.~\eqref{eq:model_bloch_hamiltonian}. Since this term is proportional to the identity, it does not change the eigenvectors, the winding number, or the edge-state localization.

The two bulk bands are obtained from the magnitude of \(q(k)\),
\begin{equation}
E_\pm(k)
=
\pm
\sqrt{
(t_1^{\mathrm{eff}})^2+
(t_2^{\mathrm{eff}})^2+
2t_1^{\mathrm{eff}}t_2^{\mathrm{eff}}\cos k
}.
\label{eq:model_single_chain_bands}
\end{equation}
The minimum band separation occurs at \(k=\pi\), giving the bulk gap
\begin{equation}
\Delta_{\mathrm{bulk}}
=
2|t_2^{\mathrm{eff}}-t_1^{\mathrm{eff}}|.
\label{eq:model_bulk_gap}
\end{equation}
Consequently, the topology can change only when this gap closes, namely when
\[
t_1^{\mathrm{eff}}=t_2^{\mathrm{eff}}.
\]

The topological invariant is the winding number of \(q(k)\) around the origin,
\begin{equation}
\nu
=
\frac{1}{2\pi}
\int_{-\pi}^{\pi}
\frac{d}{dk}\arg[q(k)]\,dk .
\label{eq:model_winding}
\end{equation}
Geometrically, \(q(k)\) traces a circle in the complex plane centered at \(t_1^{\mathrm{eff}}\) with radius \(|t_2^{\mathrm{eff}}|\). The origin is enclosed only when the intercell hopping is larger than the intracell hopping:
\[
|\nu|=
\begin{cases}
1, & |t_2^{\mathrm{eff}}|>|t_1^{\mathrm{eff}}|,\\
0, & |t_2^{\mathrm{eff}}|<|t_1^{\mathrm{eff}}|.
\end{cases}
\]
This gives the SSH bulk-boundary correspondence in the present interacting mean-field setting:
\begin{equation}
\begin{aligned}
|t_2^{\mathrm{eff}}|>|t_1^{\mathrm{eff}}|
&\Longleftrightarrow |\nu|=1
\\
&\Longleftrightarrow
\text{two open-boundary edge states}.
\end{aligned}
\label{eq:model_bulk_boundary}
\end{equation}

The open-boundary calculation is used to verify the physical consequence of the bulk invariant. In real space, the finite chain is represented by a tridiagonal matrix whose nearest-neighbor entries alternate as
\[
(H_{\mathrm{OBC}})_{j,j+1}
=
(H_{\mathrm{OBC}})_{j+1,j}
=
\begin{cases}
t_1^{\mathrm{eff}}, & j\ \mathrm{odd},\\
t_2^{\mathrm{eff}}, & j\ \mathrm{even}.
\end{cases}
\]
In the topological regime, the boundary eigenstates decay exponentially into the bulk. Their localization length is
\[
\xi=
\frac{1}{\ln|t_2^{\mathrm{eff}}/t_1^{\mathrm{eff}}|},
\]
and for \(N=2N_c\) sites the finite-size splitting of the edge doublet scales as
\[
\delta E_{\mathrm{edge}}
\propto
\left|
\frac{t_1^{\mathrm{eff}}}{t_2^{\mathrm{eff}}}
\right|^{N_c}.
\]
This splitting vanishes exponentially in the thermodynamic limit.

For numerical identification of edge-localized states, we use the probability density
\[
\rho_m(j)=|\psi_m(j)|^2
\]
of the \(m\)-th open-boundary eigenstate. The corresponding edge weight is
\[
W_{\mathrm{edge}}^{(m)}
=
\sum_{j=1}^{\ell}\rho_m(j)
+
\sum_{j=N-\ell+1}^{N}\rho_m(j),
\]
where \(\ell\) is the number of sites included near each boundary. States with large \(W_{\mathrm{edge}}^{(m)}\) and energies inside the bulk gap are identified as edge modes. The derivation of the Bloch Hamiltonian, winding number, periodic-boundary self-consistency integrals, and edge-state wavefunctions is given in Appendix~\ref{app:topology_edge}.

\subsection{Two-leg ladder extension}

We now extend the single-chain construction to a two-leg ladder in order to examine how interchain hybridization modifies the SSH edge sector. The ladder consists of two identical dimerized chains labelled by \(\lambda=1,2\). The intrachain hopping pattern is the same on both legs, while a rung hopping couples equal site indices on opposite legs. After Hartree--Fock reduction and subtraction of the uniform density shift, the ladder Hamiltonian is
\begin{align}
\hat H_{\mathrm{ladder}}^{\mathrm{MF}}
&=
\sum_{\lambda=1}^{2}
\sum_{j=1}^{N-1}
t_j^{\mathrm{eff}}
(\hat c_{\lambda,j}^\dagger\hat c_{\lambda,j+1}
+\mathrm{h.c.})
\nonumber\\
&\quad+
t_\perp^{\mathrm{eff}}
\sum_{j=1}^{N}
(\hat c_{1,j}^\dagger\hat c_{2,j}
+\mathrm{h.c.}) .
\label{eq:model_ladder_hamiltonian}
\end{align}
Here
\[
t_j^{\mathrm{eff}}
=
\begin{cases}
t_1^{\mathrm{eff}}, & j\ \mathrm{odd},\\
t_2^{\mathrm{eff}}, & j\ \mathrm{even}.
\end{cases}
\]
If a rung density interaction is retained, the rung hopping is renormalized as
\[
t_\perp^{\mathrm{eff}}
=
t_\perp-4J_z^\perp\chi_\perp .
\]
The corresponding ladder chemical-potential shift is
\[
\mu_{\mathrm{eff}}^{\mathrm{ladder}}
=
2h+2J_z^\perp .
\]
In the calculations discussed here, we focus on single-particle interchain hybridization and set \(J_z^\perp=0\). Hence
\[
t_\perp^{\mathrm{eff}}=t_\perp,
\qquad
\mu_{\mathrm{eff}}^{\mathrm{ladder}}=2h.
\]

Under open boundary conditions, the ladder has a transparent block structure. In the basis where all sites of leg 1 are listed first and all sites of leg 2 second, the single-particle Hamiltonian is
\begin{equation}
\mathcal H_{\mathrm{ladder}}^{\mathrm{OBC}}
=
\begin{pmatrix}
H_{\mathrm{OBC}} & t_\perp^{\mathrm{eff}}\mathbb I_N\\
t_\perp^{\mathrm{eff}}\mathbb I_N & H_{\mathrm{OBC}}
\end{pmatrix}.
\label{eq:model_ladder_obc_matrix}
\end{equation}
The diagonal blocks are the open-boundary SSH matrices of the two individual legs, while the off-diagonal blocks describe vertical rung hybridization.

For the bulk analysis, periodic boundary conditions are imposed along the chains. Using the same off-diagonal function \(q(k)\) as in the single-chain problem, the ladder Bloch Hamiltonian becomes
\begin{equation}
\mathcal H_{\mathrm{ladder}}(k)
=
\begin{pmatrix}
0 & q & t_\perp^{\mathrm{eff}} & 0\\
q^* & 0 & 0 & t_\perp^{\mathrm{eff}}\\
t_\perp^{\mathrm{eff}} & 0 & 0 & q\\
0 & t_\perp^{\mathrm{eff}} & q^* & 0
\end{pmatrix},
\label{eq:model_ladder_bloch}
\end{equation}
where \(q\equiv q(k)\). The rung hopping is proportional to the identity in the sublattice sector, so the matrix separates into bonding and antibonding leg combinations. Each sector is an SSH Hamiltonian shifted by \(\pm t_\perp^{\mathrm{eff}}\), giving the four-band spectrum
\begin{equation}
E_{\eta,\pm}(k)
=
\eta t_\perp^{\mathrm{eff}}
\pm
|q(k)|,
\qquad
\eta=\pm1 .
\label{eq:model_ladder_bands}
\end{equation}
When the uniform density shift is restored,
\[
E_{\eta,\pm}(k)
=
\mu_{\mathrm{eff}}^{\mathrm{ladder}}
+
\eta t_\perp^{\mathrm{eff}}
\pm
|q(k)|.
\]

Because the bonding and antibonding sectors contain the same function \(q(k)\), the rung hopping does not modify the SSH winding condition. Instead, it doubles the topological index:
\[
|\nu_{\mathrm{ladder}}|=2|\nu|.
\]
Thus, when the underlying chain is topological, the ladder supports four open-boundary edge modes. The rung hopping hybridizes the two edge states localized at the same boundary and shifts them approximately to
\[
E_{\mathrm{edge}}^\pm
\simeq
\mu_{\mathrm{eff}}^{\mathrm{ladder}}
\pm
t_\perp^{\mathrm{eff}}.
\]
The edge modes remain well defined only while this rung-induced splitting does not push them into the bulk continuum. A necessary in-gap condition is
\begin{equation}
|t_\perp^{\mathrm{eff}}|
<
|t_2^{\mathrm{eff}}-t_1^{\mathrm{eff}}|.
\label{eq:model_edge_criterion}
\end{equation}
For clearly isolated edge modes in the numerical spectra, we use the more conservative practical criterion
\begin{equation}
|t_\perp^{\mathrm{eff}}|
\lesssim
\frac12
|t_2^{\mathrm{eff}}-t_1^{\mathrm{eff}}|.
\label{eq:model_ladder_protection}
\end{equation}
The ladder construction, block diagonalization, four-band spectrum, and edge-state splitting are derived in Appendix~\ref{app:ladder}.

 
%

\subsection{Dimensionless units and experimental considerations}
\label{sec:experimental}

Throughout this work, all energies are expressed in units of the
reference hopping scale
\begin{equation}
t_0 \equiv 2\widetilde{J}_{xy}(\bar{r}),
\label{eq:t0_def}
\end{equation}
which is the chirality-renormalized transverse exchange evaluated at
the mean intermolecular separation $\bar{r}$. In these units the bare
dimerized hoppings become
\begin{equation}
\frac{t_1}{t_0}
=
\left(\frac{\bar{r}}{\bar{r}+\delta r}\right)^{\!3},
\qquad
\frac{t_2}{t_0}
=
\left(\frac{\bar{r}}{\bar{r}-\delta r}\right)^{\!3},
\label{eq:dimless_hoppings}
\end{equation}
and the Ising coupling is parametrized by $J_z/t_0$. The topological
phase structure depends only on the dimensionless ratios
$t_1^{\mathrm{eff}}/t_2^{\mathrm{eff}}$ and
$t_\perp/\Delta_{\mathrm{bulk}}$, not on the absolute energy scale.
This makes the results applicable to any physical realization of the
effective Hamiltonian in Eq.~\eqref{eq:model_bulk_fermion}, regardless of the
specific molecular species or trapping geometry.

The dimerized chain and its ladder extension can in principle be
realized using programmable optical tweezer
arrays~\cite{Ni2018,Cheuk2022}. Unlike optical lattices, where the
spatial geometry is fixed by laser interference, tweezer arrays allow
independent positioning of individual sites via spatial light
modulators~\cite{Princeton2024}. The SSH dimerization is introduced by
programming an alternating spacing
$r_j = \bar{r} - (-1)^j \delta r$ between adjacent tweezers. For a
mean spacing of $\bar{r} = 400$--$600$\,nm, the operating point
$\delta r/\bar{r} = 0.067$ used in this work corresponds to a spatial
modulation of $\pm 27$--$40$\,nm, which is within the calibration
precision of current tweezer systems. The ladder geometry is obtained
by generating a second parallel row at a transverse separation
$r_\perp > \bar{r}$. The protection criterion
$|t_\perp| < |t_2^{\mathrm{eff}} - t_1^{\mathrm{eff}}|$ translates
to a geometric bound
\begin{equation}
r_\perp
\;\gtrsim\;
\bar{r}
\left(
\frac{2\,t_0}{|t_2^{\mathrm{eff}}-t_1^{\mathrm{eff}}|}
\right)^{\!1/3},
\label{eq:rperp_bound}
\end{equation}
which for the parameters used here gives
$r_\perp \gtrsim 2.6\,\bar r \approx 1.0\,\mu\mathrm{m}$
for $\bar r = 400$\,nm.

The absolute energy scale is set by the dipolar coupling
$t_0 \propto \mu^2/\bar{r}^3$, where $\mu$ is the effective
Stark-dressed dipole moment. For a polyatomic molecule with
$\mu \approx 2$\,D at $\bar{r} = 400$\,nm, the reference hopping
is $t_0 \sim 100$\,kHz, giving a bulk topological gap of
$\Delta_{\mathrm{bulk}} \approx 0.27\,t_0 \sim 27$\,kHz.
Topological protection
against thermal excitations then requires
$T \lesssim h_{\mathrm P}\Delta_{\mathrm{bulk}}/k_B \sim 1\,\mu$K. This
threshold scales quadratically with the dipole moment: a molecule
with $\mu \approx 4$\,D yields
$\Delta_{\mathrm{bulk}} \sim 100$\,kHz, relaxing the requirement
to $T \lesssim 5\,\mu$K.

The chirality-specific features of the model require molecules with
a definite handedness. Bialkali species such as NaCs
($\mu \approx 4.6$\,D) and RbCs ($\mu \approx 1.2$\,D) are the most
advanced molecular tweezer platforms~\cite{Ni2018,Ni2021} but are
structurally achiral; they can realize the conventional SSH topology
without the stereochemical edge-state labeling predicted here. A
route to chiral species is suggested by recent progress in laser
cooling of polyatomic symmetric-top molecules~\cite{Doyle2020}. In
particular, CaOCH$_3$ has been laser-cooled to the sub-millikelvin
regime. Although CaOCH$_3$ itself is achiral ($C_{3v}$ symmetry),
substitution of the methyl hydrogens with distinct groups — forming,
for example, a $-$CHDT or $-$CFClH center — would create a chiral
derivative while preserving the calcium cycling center needed for
optical trapping~\cite{Doyle2020}. An alternating L--R lattice could
then be assembled using enantiomer-selective loading or microwave
sorting prior to trapping. Once realized, the chirality-induced DM
interaction would amplify the effective hopping to
$\widetilde{J}_{xy} = \sqrt{J_{xy}^2 + D^2}$, providing a structural
enhancement of the bulk topological gap beyond what an achiral chain
of the same dipole moment can achieve.

\section{Results and Discussion}
\label{sec:results}

We organize the results around three complementary diagnostics applied to both the single-chain and two-leg ladder geometries: bulk energy spectra under periodic boundary conditions, the complex-plane winding of the effective hopping function, and finite-size spectra with real-space wavefunction analysis under open boundary conditions. Together these diagnostics establish the trivial--topological phase structure of the model and illuminate the dual role of molecular chirality: as the microscopic source of the chirality-renormalized hopping scale and as the quantum label that distinguishes the left- and right-boundary edge states.

\subsection{Bulk spectra under periodic boundary conditions}

\begin{figure*}[!t]
    \centering
    \fbox{\includegraphics[width=0.94\textwidth]{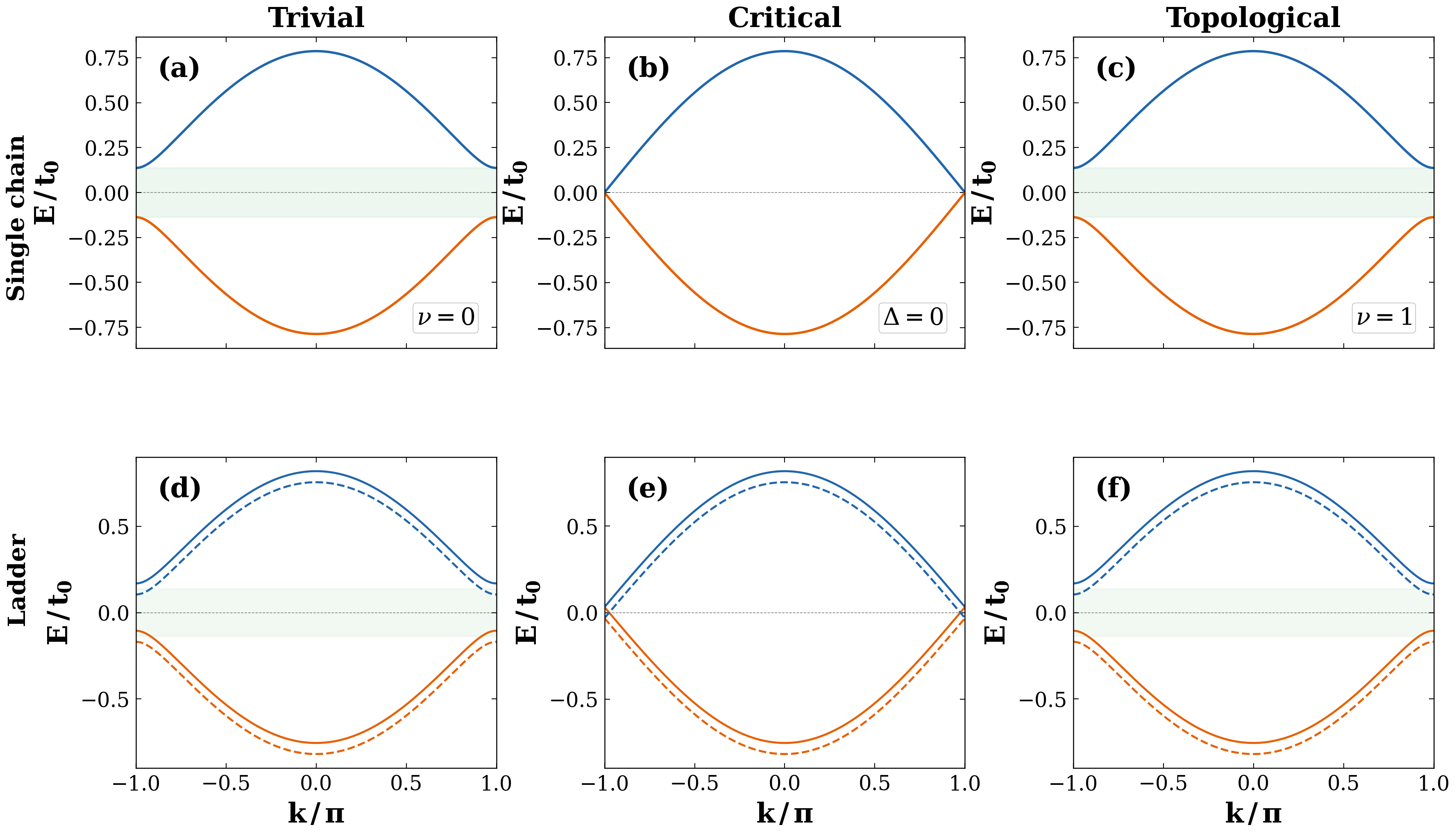}}
\caption{Bulk energy spectra $E_\pm(k)$ under periodic boundary
conditions for the self-consistently treated dimerized chiral molecular
model. The reference energy scale is $t_0 = 2\widetilde{J}_{xy}(\bar{r})$;
the Ising coupling is $J_z/t_0 = -0.190$ and the system is at half
filling. The green-shaded band marks the bulk gap
$\Delta_{\mathrm{bulk}} = 2|t_2^{\mathrm{eff}} - t_1^{\mathrm{eff}}|$
in all panels; the horizontal dotted line marks $\varepsilon_0 = 0$.
\emph{Single chain} (upper row): the blue $E_+(k)$ and orange $E_-(k)$
bands are determined by the mean-field effective hoppings
$t_1^{\mathrm{eff}}$ and $t_2^{\mathrm{eff}}$ in the
(a)~trivial ($\delta r/\bar{r} = -0.067$, $\nu = 0$),
(b)~critical ($\delta r = 0$, $\Delta_{\mathrm{bulk}} = 0$), and
(c)~topological ($\delta r/\bar{r} = +0.067$, $\nu = 1$) regimes.
\emph{Ladder} (lower row): corresponding two-leg ladder spectra with
rung coupling $t_\perp/t_0 = 0.032$ for the
(d)~trivial, (e)~critical, and (f)~topological regimes.
The interchain hybridization splits each SSH band into a bonding
sector (solid lines) and an antibonding sector (dashed lines),
producing the four-band structure
$E = \varepsilon_0 \pm t_\perp \pm |q(k)|$.
The same trivial--critical--topological sequence persists in both
rows, confirming that the ladder inherits its bulk topology from
the underlying dimerized chains.}
    \label{fig:pbc_spectrum}
\end{figure*}

The single-chain bulk energies follow from Eqs.~\eqref{eq:model_single_chain_bands} and \eqref{eq:model_bulk_gap}, and the resulting spectra are shown in Fig.~\ref{fig:pbc_spectrum}. Three regimes are visible in the upper row. In the trivial phase ($\delta r<0$, $t_1>t_2$), the mean-field converges to $t_1^{\mathrm{eff}}>t_2^{\mathrm{eff}}$ and the chain is continuously connected to the ordinary SSH trivial insulator. At the critical point ($\delta r=0$), the self-consistent bond orders satisfy $\chi_1=\chi_2$, the effective dimerization vanishes, and the gap closes as required by $t_1^{\mathrm{eff}}=t_2^{\mathrm{eff}}$. In the topological phase ($\delta r>0$, $t_2>t_1$), the gap reopens about a distinct bulk configuration with $t_2^{\mathrm{eff}}>t_1^{\mathrm{eff}}$.

\paragraph*{Role of molecular chirality in the bulk spectrum.}
A fundamental ingredient that distinguishes this platform from a conventional electronic SSH chain is the Dzyaloshinskii--Moriya (DM) interaction $D_j$ generated by the molecular chirality of the alternating L-R array. Although $D_j$ is fully removed from the open-chain bulk by the local gauge rotation described in Appendix~\ref{app:gauge_jw}, it enters the effective hopping through the chirality-renormalized exchange scale $\tilde{J}_{xy,j}$ defined in Eq.~\eqref{eq:results_tildeJ}:
\begin{equation}
\tilde{J}_{xy,j} = \sqrt{J_{xy,j}^2 + D_j^2} \geq |J_{xy,j}|.
\label{eq:results_tildeJ}
\end{equation}
Molecular chirality therefore \emph{amplifies} both bare hoppings $t_1$ and $t_2$ beyond the value that exchange coupling alone would produce. At the operating point used in this work, the DM term raises the magnitude of the transverse exchange from
$|J_{xy}|$ to
$\tilde{J}_{xy} = \sqrt{J_{xy}^2 + D^2} \approx 1.054\,|J_{xy}|$, a
$5.4\%$ enhancement. This chirality-induced amplification is a robust feature of the L-R alternating architecture: any chain in which molecular handedness alternates bond by bond will exhibit this enhancement, making the chiral molecular platform intrinsically favorable for realizing large-gap topological phases compared to an achiral chain with the same exchange magnitude.

\paragraph*{Interaction-driven renormalization of the phase boundary.}
The Ising interaction $J_z$ modifies the phase boundary through the self-consistent bond orders given in Eq.~\eqref{eq:model_effective_hoppings}. In the dimerized phase the two bond orders are unequal: the stronger bond (whichever of $t_1$ or $t_2$ dominates) develops the larger magnitude of bond coherence. Because $J_z<0$ at the operating point and the half-filled SSH bond orders are negative in the convention used here, the Fock correction $-4J_z\chi_j<0$ reduces both effective hoppings from their bare values, with the reduction being \emph{larger} for the stronger bond. Consequently, the effective dimerization ratio $t_2^{\mathrm{eff}}/t_1^{\mathrm{eff}}$ is compressed toward unity relative to the bare ratio $t_2/t_1$: attractive interactions weaken the effective dimerization and move the system toward the critical point. In principle, a sufficiently strong attractive $J_z$ can drive an interaction-induced topological phase transition even without changing $\delta r$, while repulsive interactions ($J_z>0$) would enhance the effective dimerization and push the gapped phases further apart. A further consequence of the bond-order asymmetry is that attractive interactions reduce the effective dimerization and hence reduce the bulk gap relative to the bare SSH value. In finite open chains, small termination-dependent differences can still appear in the edge-state splitting, but the bulk topological criterion remains controlled by $|t_2^{\mathrm{eff}}-t_1^{\mathrm{eff}}|$.

The lower row of Fig.~\ref{fig:pbc_spectrum} shows the bulk spectra for the two-leg ladder. The rung hopping $t_\perp$ hybridizes the two chains and produces four bands according to Eq.~\eqref{eq:model_ladder_bands},
\begin{equation}
E_{s_1 s_2}(k)=\varepsilon_0 + s_1 t_\perp + s_2\left|q(k)\right|, \quad s_1,s_2\in\{+1,-1\}.
\label{eq:results_ladder_bands}
\end{equation}
The bonding ($s_1=+1$) and antibonding ($s_1=-1$) sectors are clearly resolved in Fig.~\ref{fig:pbc_spectrum}(d)--(f). The same trivial--critical--topological sequence is visible in both sectors, confirming that the ladder inherits its bulk topology from the underlying dimerized chains.

\subsection{Complex-plane winding and bulk topology}
\label{sec:winding}

\begin{figure*}[!t]
    \centering
    \fbox{\includegraphics[width=0.98\textwidth]{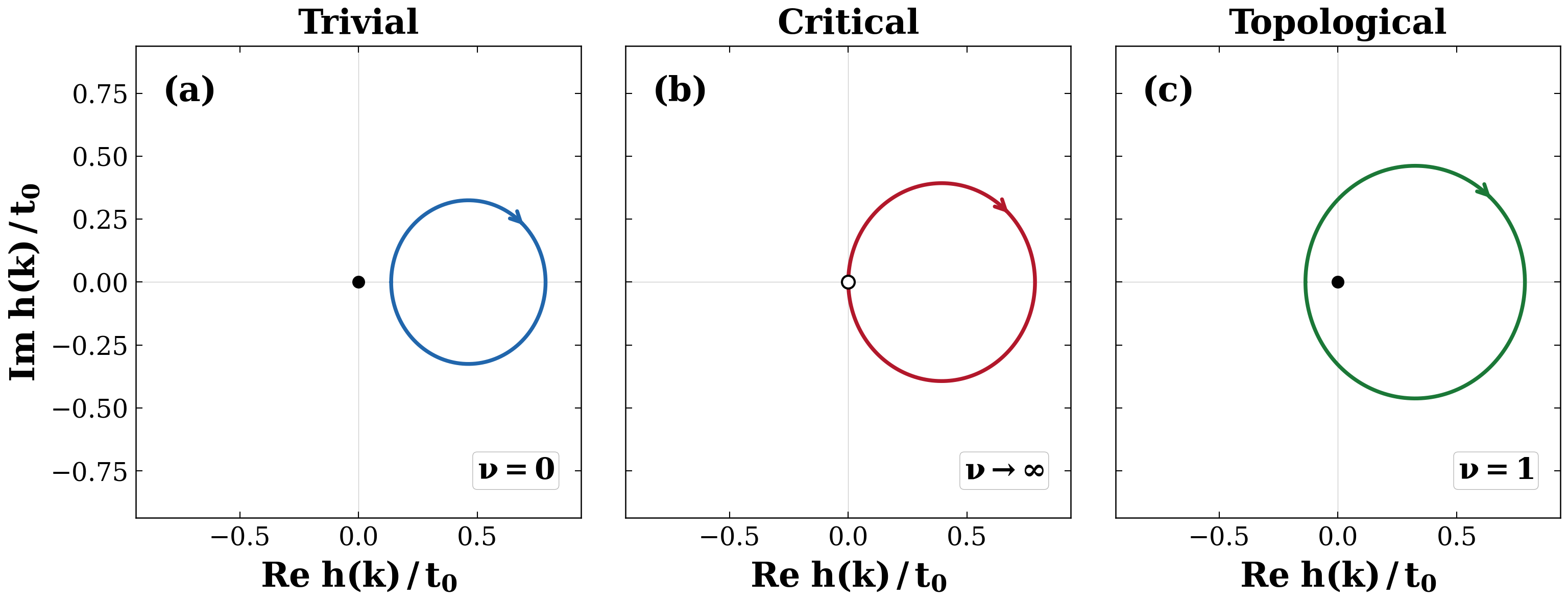}}
    \caption{Complex-plane winding trajectories of
$q(k) = t_1^{\mathrm{eff}} + t_2^{\mathrm{eff}}e^{-ik}$
as $k$ traverses the full Brillouin zone $[0,2\pi)$, for the
(a)~trivial, (b)~critical, and (c)~topological phases.
The coloured loop is the trajectory of $q(k)$; the filled black
dot marks the origin $(0,0)$ in panels~(a) and~(c), while the
open circle in panel~(b) indicates that the loop passes exactly
through the origin at the critical point.
All three panels share the same axis scale, permitting direct
comparison: in the trivial phase the loop lies entirely to the
right of the origin ($\nu = 0$); at the critical point the loop
passes through the origin, so the winding number is undefined
($\Delta_{\mathrm{bulk}} = 0$); in the topological phase the loop
encircles the origin once ($\nu = 1$).
The loop radius equals $t_2^{\mathrm{eff}}$ and the
center-to-origin distance equals $t_1^{\mathrm{eff}}$; the
condition $t_2^{\mathrm{eff}} > t_1^{\mathrm{eff}}$ for
enclosure directly connects the geometric picture to the
bulk--boundary correspondence.
In the ladder geometry, each chain contributes independently to
the winding, giving $\nu_{\mathrm{ladder}} = 2\nu_{1\mathrm{D}}$
and predicting four OBC edge modes in the topological phase
(see \linebreak[2] Appendix~\ref{app:ladder})}
    \label{fig:winding}
\end{figure*}

The bulk-topological distinction between the two gapped phases is not encoded in the eigenvalue spectra alone, since interchanging $t_1^{\mathrm{eff}}$ and $t_2^{\mathrm{eff}}$ leaves $|q(k)|$ unchanged \cite{SSH1979,Asboth2016}. The appropriate bulk invariant is the winding number of the complex hopping function $q(k)$ defined in Eq.~\eqref{eq:model_hk}, given by Eq.~\eqref{eq:model_winding}. Geometrically, $q(k)$ traces a circle in the complex plane centered at $(t_1^{\mathrm{eff}},0)$ with radius $t_2^{\mathrm{eff}}$. The origin is enclosed ($\nu=1$) when $t_2^{\mathrm{eff}}>t_1^{\mathrm{eff}}$ and excluded ($\nu=0$) when $t_1^{\mathrm{eff}}>t_2^{\mathrm{eff}}$; at the critical point the circle passes through the origin and $\nu$ is ill-defined.

Fig.~\ref{fig:winding} shows the winding trajectories for the three regimes. In Fig.~\ref{fig:winding}(a), the trivial-phase loop lies entirely to the right of the origin. In Fig.~\ref{fig:winding}(b), the critical loop touches the origin. In Fig.~\ref{fig:winding}(c), the topological loop encircles the origin once. All panels are plotted on a common axis scale, making the geometric distinction immediately apparent.

The molecular chirality influences this picture in a subtle but important way. Before the gauge rotation, the complex bond coupling $J_{xy,j}+iD_j$ adds an imaginary component to the hopping, rotating $q(k)$ in the complex plane. The gauge transformation absorbs this phase into $\tilde{J}_{xy,j}$, enlarging the radius of the loop (since the radius equals $t_2^{\mathrm{eff}}\propto\tilde{J}_{xy}>|J_{xy}|$) while preserving the topological classification. Thus molecular chirality not only amplifies the bulk gap but also enlarges the geometric loop in the winding diagram, providing a broader margin between the loop and the origin in the topological phase and making the winding number more robust against perturbations.

For the ladder, the rung coupling $t_\perp$ enters as a diagonal shift and does not affect the winding of the off-diagonal function. Consequently (see Appendix~\ref{app:ladder}), $\nu_{\mathrm{ladder}}=2\nu_{1\mathrm{D}}$, taking the value $2$ in the topological regime. This doubled winding predicts four OBC edge modes, as confirmed in the open-boundary analysis below.

\subsection{Open-boundary spectra and edge-sector evolution}

\begin{figure*}[!t]
    \centering
    \fbox{\includegraphics[width=0.97\textwidth]{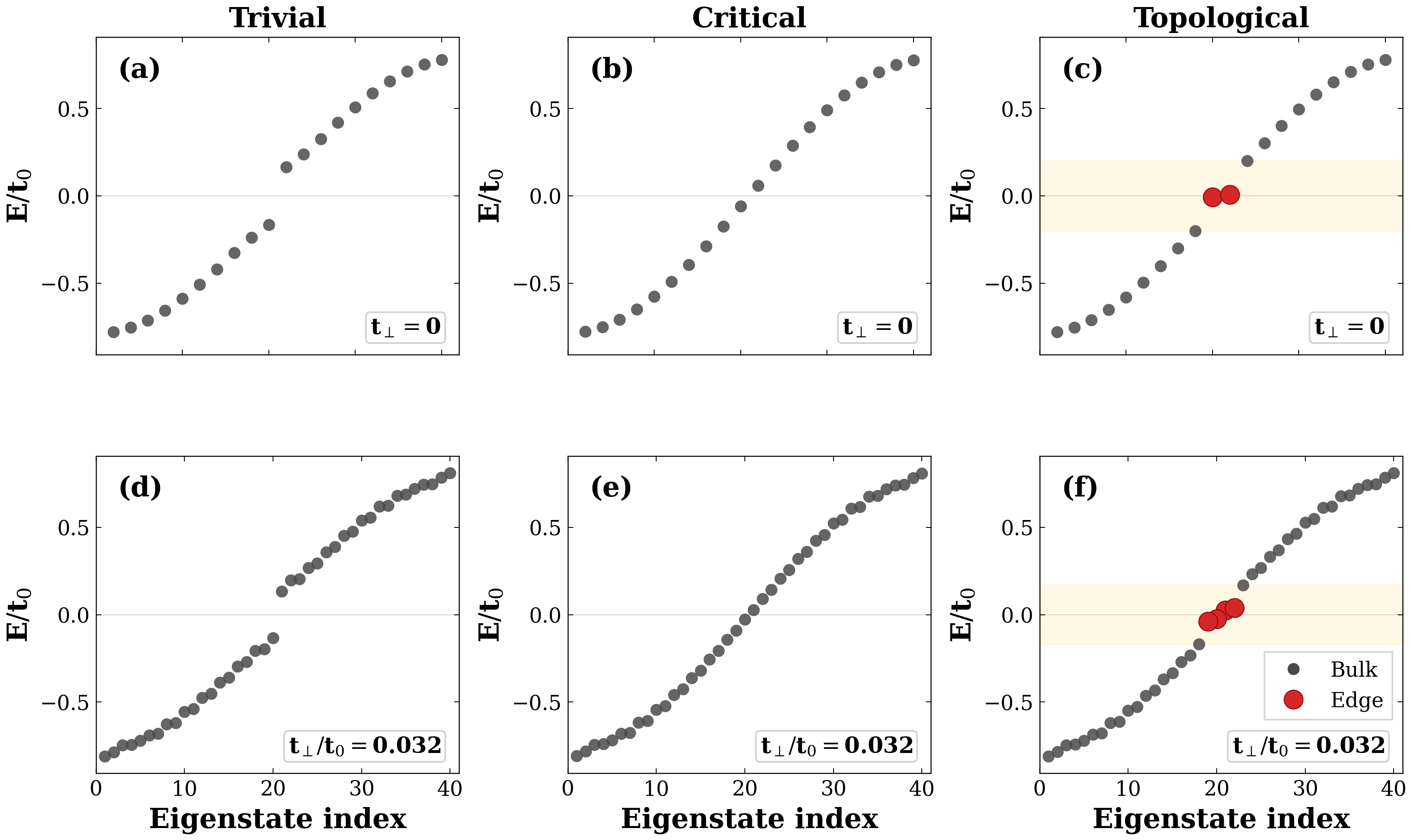}}
    \caption{Open-boundary energy spectra as a function of eigenstate
index for a system of $N = 20$ sites per chain ($N_c = 10$ unit
cells, 40 total eigenstates). Gray circles denote bulk states;
red circles highlight in-gap edge states (see legend in
panel~(f)). The solid horizontal line marks $\varepsilon_0 = 0$;
the pale-yellow band in panels~(c) and~(f) indicates the spectral
gap region containing the edge states.
Upper row~(a)--(c): two decoupled dimerized chains ($t_\perp = 0$)
in the trivial, critical, and topological regimes.
In the trivial regime~(a) no isolated boundary sector appears.
At the critical point~(b) the bulk gap closes and no spectrally
isolated edge sector exists.
In the topological regime~(c) two red points appear inside the
gap near $\varepsilon_0$; each point represents a near-degenerate
pair of edge states from the two identical decoupled chains, whose
eigenvalues overlap exactly, giving a total of four edge modes.
Their small residual splitting reflects finite-size hybridization
scaling as $(t_1^{\mathrm{eff}}/t_2^{\mathrm{eff}})^{N_c}$.
Lower row~(d)--(f): coupled two-leg ladder with
$t_\perp/t_0 = 0.032$.
The trivial~(d) and critical~(e) ladder phases show no isolated
boundary modes.
In the topological ladder regime~(f) the rung hopping lifts the
degeneracy of the four edge modes, splitting them into two
distinct pairs visible as four separate red points inside the
gap; the in-gap states remain spectrally isolated provided
$t_\perp < |t_2^{\mathrm{eff}} - t_1^{\mathrm{eff}}|$
(Eq.~\eqref{eq:model_edge_criterion}).}
    \label{fig:obc_spectrum}
\end{figure*}

The bulk--boundary correspondence prediction of the winding analysis---one pair of OBC edge states per chain in the topological regime---is directly confirmed by the finite-size spectra shown in Fig.~\ref{fig:obc_spectrum}. We use these spectra to build the complete story connecting ordinary SSH edge states to the chiral molecular edge states specific to this platform.

\paragraph*{Ordinary SSH edge states.}
In the simplest picture, the edge states of a dimerized chain are the zero modes of the SSH Hamiltonian, protected by the sublattice (chiral) symmetry $\{H,\Gamma\}=0$, where $\Gamma=\mathrm{diag}(+1,-1,+1,-1,\ldots)$ alternates $+1$ and $-1$ on odd and even sites. This symmetry pins the edge-state energies to $\varepsilon_0$ in the thermodynamic limit and classifies the SSH chain in symmetry class BDI~\cite{SSH1979,Heeger1988,Asboth2016}. The upper row of Fig.~\ref{fig:obc_spectrum} shows the decoupled limit ($t_\perp=0$), where each chain independently realizes the trivial, critical, or topological phase. In Fig.~\ref{fig:obc_spectrum}(c), four in-gap states appear in the topological regime: two from each chain, corresponding to one mode localized near each physical boundary of each leg.

\paragraph*{Chiral edge states: the molecular dimension.}
The chiral molecular platform adds a physically significant new dimension to this picture. In the alternating L-R chain, odd sites ($j=1,3,5,\ldots$) are associated with left-handed (L) molecules and even sites ($j=2,4,6,\ldots$) with right-handed (R) molecules. In the topological phase, with $t_2>t_1$, the intercell bond $t_2$ is dominant, leaving the outermost sites essentially \emph{unpassivated}: the left-boundary edge state is exponentially concentrated on site $j=1$, which carries the chirality of the \emph{left-handed} molecule, while the right-boundary edge state is concentrated on site $j=N$, which carries the chirality of the \emph{right-handed} molecule. The two in-gap modes therefore carry opposite molecular chirality --- they are \emph{chiral edge states} in the molecular sense, directly encoding the handedness of the host molecule at each boundary.

This is a qualitatively new feature beyond ordinary SSH edge states, in which boundary sites are indistinguishable lattice points. In the chiral molecular chain the edge modes inherit the stereochemical identity of the molecule on which they are predominantly localized. The left-chiral and right-chiral edge states are related by a combined operation of spatial inversion and chirality reversal (L\,$\leftrightarrow$\,R), a symmetry broken by the explicit alternation of molecular handedness in the bulk. This makes the two states physically distinct even before any external perturbation, and their different molecular character makes them separable in principle by chiroptical spectroscopies such as circular dichroism or vibrational optical activity, opening a route to chirality-selective boundary-state detection.

\paragraph*{Effect of the longitudinal interaction on edge-state energies.}
The Ising coupling $J_z$ influences the edge states both by renormalizing the bulk gap and by modifying the finite-size splitting. Because the self-consistent bond orders satisfy $\chi_{\mathrm{strong}}\neq\chi_{\mathrm{weak}}$, a more attractive $J_z$ compresses $t_1^{\mathrm{eff}}/t_2^{\mathrm{eff}}$ toward unity, increasing the finite-size energy splitting, which scales as
\begin{equation}
\delta E_{\mathrm{edge}}\propto\left(\frac{t_1^{\mathrm{eff}}}{t_2^{\mathrm{eff}}}\right)^{N_c}.
\label{eq:results_edge_split}
\end{equation}
Conversely, reducing $|J_z|$ deepens the topological phase and makes the edge states more nearly degenerate. The edge-state splitting thus provides a sensitive indirect probe of the interaction strength accessible through spectroscopy: by measuring the energy difference between the two near-midgap states as a function of intermolecular separation (which controls $J_z\propto r^{-3}$), one can map out the interaction-renormalized phase diagram.

\paragraph*{Spatial localization of the edge states.}
The real-space structure of the edge states is controlled by the effective dimerization ratio. The localization length is
\begin{equation}
\xi = \frac{a}{\left|\ln\!\left(t_1^{\mathrm{eff}}/t_2^{\mathrm{eff}}\right)\right|},
\label{eq:results_xi}
\end{equation}
where $a$ is the lattice constant. For the operating parameters at $\delta r/\bar r = +0.067$, we estimate
$t_1^{\mathrm{eff}}/t_2^{\mathrm{eff}} \approx 0.73$, giving $\xi\approx 3.1\,a$ and $\delta E_{\mathrm{edge}}\propto (0.73)^{10}\approx 0.04$ for $N_c=10$ unit cells. The edge states are therefore well-separated from each other in a chain of twenty sites, with a splitting an order of magnitude smaller than the bulk gap.

\paragraph*{Ladder edge states and rung-induced splitting.}
When the two chains are coupled into a ladder by a rung hopping $t_\perp$, as shown in Fig.~\ref{fig:obc_spectrum}(d)--(f), the boundary modes hybridize across the rungs. The four-fold edge sector splits into two pairs at energies approximately $\varepsilon_0\pm t_\perp$. The in-gap states remain spectrally isolated as long as the necessary in-gap condition in Eq.~\eqref{eq:model_edge_criterion} is satisfied, which holds in Fig.~\ref{fig:obc_spectrum}(f) with $t_\perp/t_0 = 0.032$ and
$|t_2^{\mathrm{eff}} - t_1^{\mathrm{eff}}|/t_0 \approx 0.12$. Importantly, since the rung hopping connects corresponding sites of the two identical chains without mixing L and R molecules at corresponding rungs, the molecular chirality label of each edge state is preserved under rung coupling, providing a robust chiral quantum number for the ladder boundary modes.

\subsection{Real-space structure of the low-energy states}
\label{sec:wavefunctions}   

\begin{figure*}[!t]
    \centering
    \fbox{\includegraphics[width=0.97\textwidth]{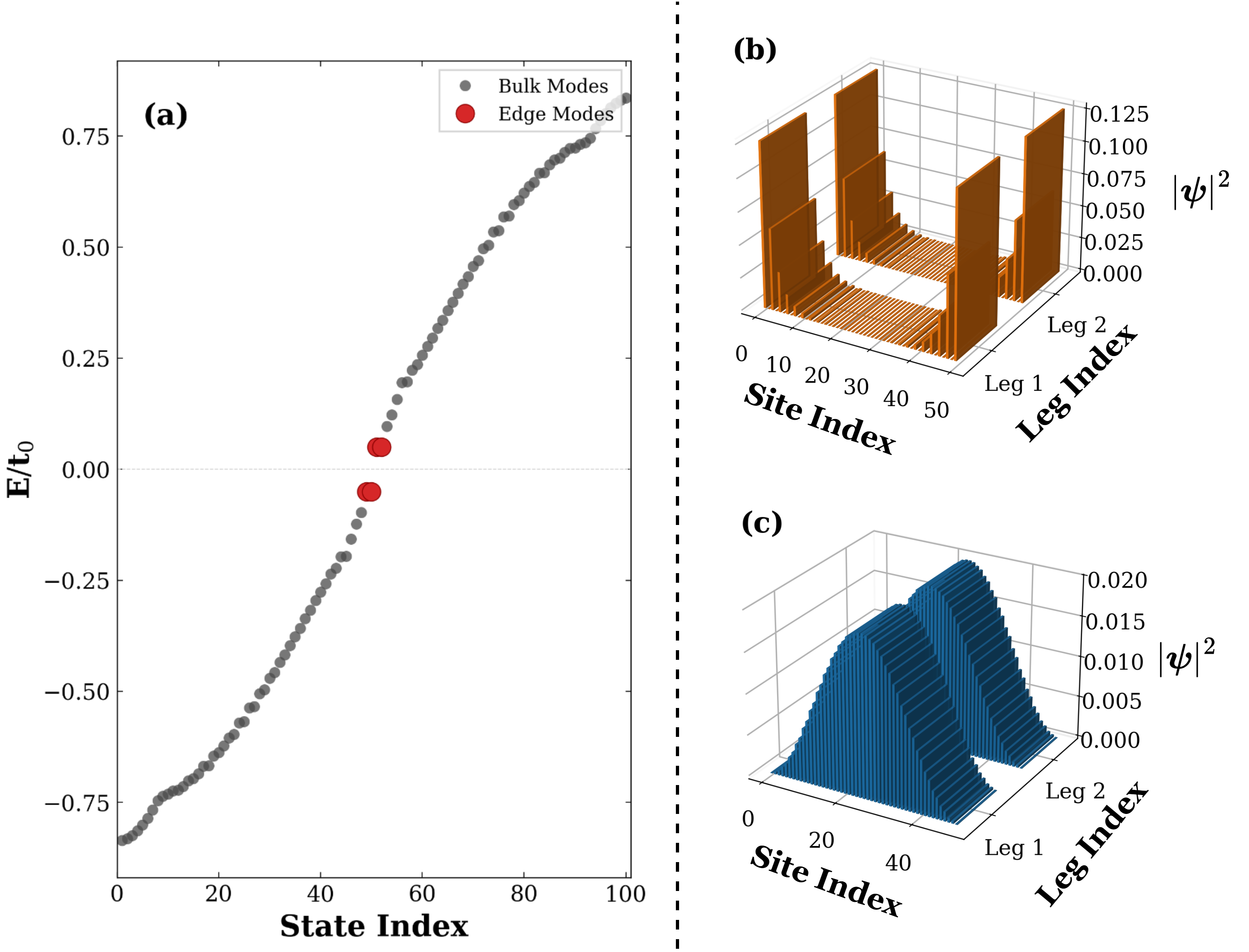}}
 \caption{Ladder boundary modes and real-space probability densities
in the topological regime
($t_1^{\mathrm{eff}}/t_0 = 0.325$, $t_2^{\mathrm{eff}}/t_0 = 0.462$,
$t_\perp/t_0 = 0.050$, $N_c = 25$ unit cells per leg,
100 total eigenstates).
(a)~Open-boundary energy spectrum as a function of state index.
Gray circles denote bulk modes; red circles identify the rung-split
in-gap edge sector closest to $E = 0$. Each visible red marker
represents a near-degenerate boundary pair, giving four ladder edge
modes in total, spectrally isolated from the bulk continuum inside
the topological gap.
(b)~Site-resolved probability density $|\psi_j|^2$ of one edge
mode, displayed separately for Leg~1 and Leg~2. The weight is
strongly concentrated at the first and last sites of both legs
and decays exponentially toward the chain interior, confirming
boundary localization on both legs of the ladder. The weight
appears at both ends because, for a finite system, the left- and
right-localized modes hybridize weakly through the bulk to form
symmetric and antisymmetric combinations; in the thermodynamic
limit these separate into independent boundary-localized states.
(c)~Site-resolved probability density of a representative bulk
mode. The weight forms a broad, smooth envelope distributed
across the chain interior with no enhancement at the boundaries,
in sharp contrast to panel~(b). The comparison between
panels~(b) and~(c) provides direct real-space confirmation that
the in-gap states identified in panel~(a) are genuinely
boundary-localized edge modes and not bulk states near the gap
edge.}
    \label{fig:wavefunctions}
\end{figure*}
The open-boundary spectra of the preceding section establish the spectral position of the edge states; we now examine their real-space structure to confirm their genuinely boundary-localized character. Fig.~\ref{fig:wavefunctions} displays the open-boundary spectrum and representative probability densities for the topological two-leg ladder.

Fig.~\ref{fig:wavefunctions}(a) shows the finite-size spectrum, where the red points identify the in-gap boundary sector separated from the bulk continuum. Fig.~\ref{fig:wavefunctions}(b) shows the probability density of a representative edge mode. The weight is strongly concentrated near the first and last sites of both legs and decays toward the interior, confirming boundary localization in the ladder geometry. Because the system is finite, the left- and right-localized states weakly hybridize through the bulk, so the exact eigenstates carry weight at both ends; in the thermodynamic limit these combinations separate into independent boundary modes.

Fig.~\ref{fig:wavefunctions}(c) shows a representative bulk state. In contrast to the edge mode in Fig.~\ref{fig:wavefunctions}(b), its probability density is distributed through the chain interior and shows no boundary enhancement. The comparison between Fig.~\ref{fig:wavefunctions}(b) and Fig.~\ref{fig:wavefunctions}(c) therefore provides direct real-space confirmation that the in-gap states identified in Fig.~\ref{fig:wavefunctions}(a) are genuinely boundary-localized ladder edge modes rather than ordinary bulk states near the gap edge.

\subsection{Edge-state evolution and robustness against rung coupling}
\label{sec:robustness}

\begin{figure*}[!t]
    \centering
    \fbox{\includegraphics[width=0.98\textwidth]{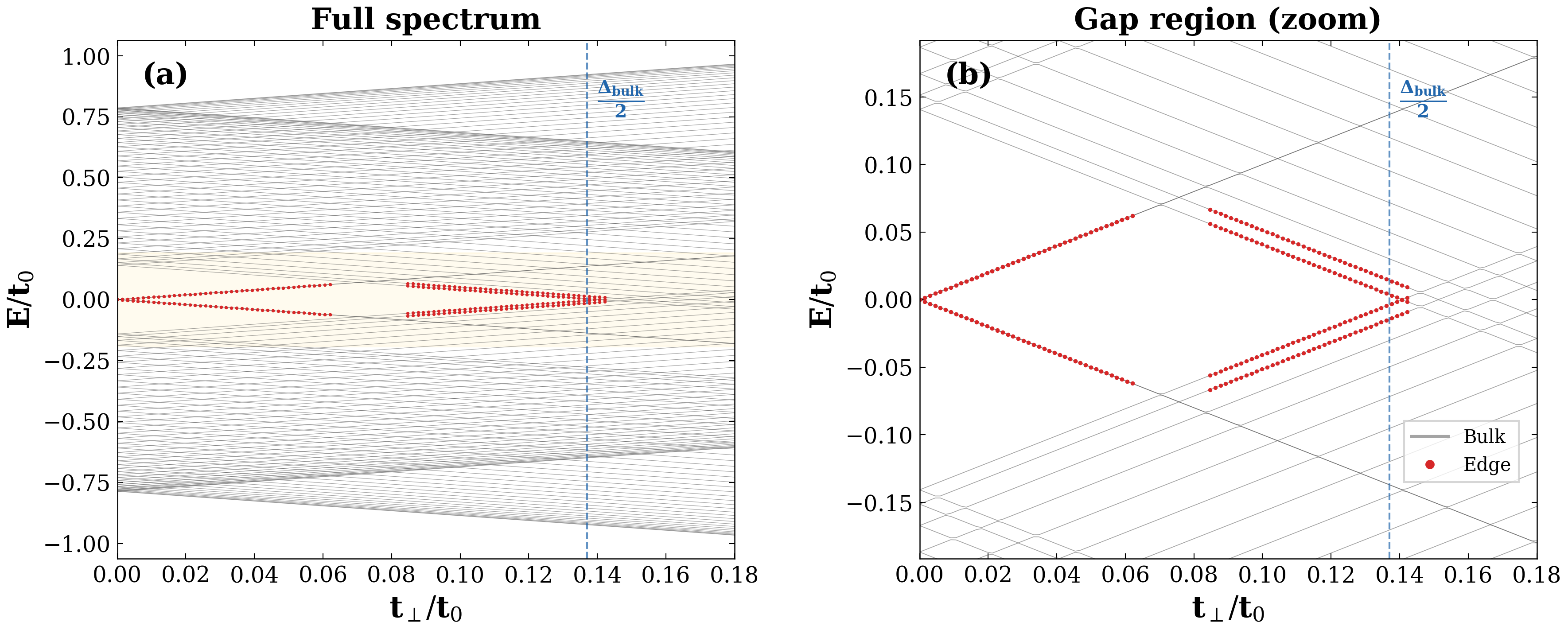}}
\caption{Evolution of the open-boundary energy spectrum of the
topological two-leg ladder as a function of the rung coupling
$t_\perp/t_0$, with $t_1^{\mathrm{eff}}/t_0 = 0.325$,
$t_2^{\mathrm{eff}}/t_0 = 0.462$, and $N_c = 40$ unit cells per
leg. Grey lines trace the bulk eigenvalues and red filled circles
mark the four edge states (see legend in panel~(b)).
(a)~Full spectrum. The yellow-shaded band marks the bulk gap
region. The edge states remain clustered near $E = 0$ for small
$t_\perp$ and merge into the bulk continuum beyond the protection
threshold (dashed blue vertical line).
(b)~Zoom into the gap region. At $t_\perp = 0$ the four edge
states are nearly degenerate at $E = 0$. As $t_\perp$ increases
the rung hybridization splits them into two symmetric pairs,
which fan outward in a characteristic diamond pattern, reach
maximum separation near $t_\perp/t_0 \approx 0.07$, and then
converge back toward $E = 0$ as the threshold is approached.
The dashed blue line marks the protection threshold
$t_\perp/t_0 = \Delta_{\mathrm{bulk}}/2 =
|t_2^{\mathrm{eff}} - t_1^{\mathrm{eff}}|/t_0 \approx 0.137$,
beyond which the rung-induced splitting exceeds the half-gap and
the edge modes merge into the bulk continuum, destroying the
spectrally isolated in-gap sector.}
    \label{fig:robustness}
\end{figure*}
The preceding subsections established the phase structure, winding, and
boundary localization at fixed operating parameters. A natural follow-up
question is how the ladder edge-state sector responds as the rung coupling
$t_\perp$ is varied continuously from zero into the regime where the
interchain hybridization overwhelms the dimerization gap.
Fig.~\ref{fig:robustness} addresses this by showing the full
open-boundary spectrum of the topological ladder as a function of
$t_\perp$, providing a continuous picture of the edge-state evolution
rather than discrete snapshots at selected coupling values.

Fig.~\ref{fig:robustness}(a) displays the complete eigenvalue fan as $t_\perp$ is swept from
zero to $0.17\,t_0$. At $t_\perp = 0$ the spectrum consists of two
independent SSH chains, each contributing one nearly degenerate edge-state
pair near $E = 0$, for a total of four midgap modes. As $t_\perp$
increases, the rung hybridization splits each bonding--antibonding sector
of the bulk bands, broadening the overall bandwidth while progressively
narrowing the gap. The four edge states, shown in red, fan outward from
the midgap cluster and eventually reach the bulk band edges, at which
point they lose their spectral isolation and merge into the continuum.

Fig.~\ref{fig:robustness}(b) shows the gap region in detail. Three regimes are clearly
visible. For small $t_\perp$, the four edge states are well separated
from the nearest bulk levels by a wide spectral gap, and the system is in
the protected regime. As $t_\perp$ approaches the threshold
$\Delta_{\mathrm{bulk}}/2 = |t_2^{\mathrm{eff}} - t_1^{\mathrm{eff}}|
\approx 0.137\,t_0$ (dashed blue line), the outermost edge-state pair
converges on the bulk band edges and the spectral isolation becomes
marginal. Beyond this threshold, the edge modes are absorbed into the bulk
and no isolated in-gap feature survives. The necessary in-gap condition in Eq.~\eqref{eq:model_edge_criterion} is confirmed quantitatively by the sweep: the edge states remain inside the gap for the entire range $t_\perp < \Delta_{\mathrm{bulk}}/2$ and disappear beyond it.

A notable feature of the sweep is that the transition from protected to
unprotected is smooth rather than abrupt. The spectral gap between the
outermost edge mode and the nearest bulk level shrinks continuously as
$t_\perp$ increases, so there is no sharp phase transition associated
with the loss of edge-state isolation. Instead, the protection degrades
gradually, which means that in a finite-temperature or disordered
experiment the effective protection window is somewhat narrower than the
zero-temperature bound of Eq.~\eqref{eq:model_edge_criterion}. Nonetheless, the
physical operating point of the chiral molecular ladder lies well inside
the protected regime, with edge states separated from the bulk by a gap
that remains a substantial fraction of $\Delta_{\mathrm{bulk}}$ for
$t_\perp$ values accessible in current molecular array geometries.

Taken together, the continuous sweep of Fig.~\ref{fig:robustness}
completes the characterization of the ladder edge-state physics. The
boundary modes are topologically mandated by the doubled bulk winding discussed in Sec.~III.B, spatially localized at the chain ends (Fig.~\ref{fig:wavefunctions}), and spectrally protected against rung
hybridization provided $t_\perp$ remains below the half-gap threshold.
The sweep format makes the robustness window and its boundaries visible
at a glance, without the need for discrete parameter labels or diagnostic
annotations.
%

\section{Conclusions}
\label{sec:conclusion}

In this work, we have developed a dimerized topological extension of
an effective chiral molecular spin platform and analyzed its bulk and
boundary properties in both single-chain and two-leg ladder geometries.
Starting from the low-energy interacting model generated by
Stark-dressed chiral molecules, we introduced bond dimerization and
showed that the resulting system naturally acquires the structure of an
SSH-like lattice with interaction-renormalized couplings, providing a
direct route from molecular chirality and dipolar interactions to
one-dimensional topological phases.

The bulk analysis under periodic boundary conditions revealed the
expected sequence of trivial, critical, and topological regimes. The
transition is marked by the closing of the dimerization gap, while the
distinction between the two gapped phases is established by the winding
of the effective hopping function in the complex plane. Although the
trivial and topological phases display similar dispersions at the level
of eigenvalues, their winding properties clearly distinguish the two
sectors. In the ladder geometry, the same basic topological structure
persists while the spectrum is enriched by interchain hybridization into
a characteristic four-band bulk structure.

Under open boundary conditions, the bulk-topological distinction becomes
directly visible in the finite-size spectra. In the trivial regime no
isolated in-gap boundary sector is present, whereas in the topological
regime low-energy boundary states emerge. In the decoupled-chain limit
the topological ladder supports four low-energy modes inherited from the
two independent topological legs; once the chains are coupled, these
modes are split by rung hybridization into a richer low-energy edge
sector. The corresponding real-space probability densities confirm the
boundary character of these states: in the topological regime the
low-energy eigenstates acquire strong weight at the system boundaries
on both legs of the ladder, while in the trivial regime the same states
remain distributed across the bulk interior. Together with the bulk
spectra and winding analysis, these results complete the bulk--boundary
picture of the dimerized chiral molecular chain.

The broader significance of the present work is that it extends the role
of chiral molecular arrays beyond their original interacting-spin
interpretation. The same platform that supports chirality-induced
exchange and effective XXZ-type interactions can also serve as a natural
setting for engineered topological boundary physics once dimerization is
introduced. A feature that distinguishes this platform from conventional
SSH implementations is that the two boundary modes carry opposite
molecular chirality: the left edge state is localized predominantly on
an L-molecule site and the right edge state on an R-molecule site. This
stereochemical labeling is a direct consequence of the alternating
handedness of the molecular array and has no analogue in achiral
tight-binding chains. It raises the possibility of chiroptical
addressing of individual boundary modes and of using the edge-state
spectrum as a local probe of chiral molecular environments, although
both applications will require dedicated analysis beyond the scope of
the present work.

Several natural theory extensions follow from the present analysis. It
would be valuable to examine the robustness of the dimerized topological
phases beyond mean-field theory using exact diagonalization,
density-matrix-renormalization-group, matrix-product-state, and
Green's-function-based many-body diagnostics. This is especially relevant
because interactions can modify one-dimensional topological classifications
and boundary-state structure in ways that are not captured by a purely
single-particle description~\cite{FidkowskiKitaev2011,TangWen2012,Manmana2012,DiSalvo2024}.
It would also be interesting to study wider ladders, higher-dimensional
generalizations, and the interplay between stronger interactions and
topological boundary structure, in order to clarify how far the present
dimerized phases persist beyond the effective single-particle description.

The results of this work are expressed throughout in dimensionless units
of the reference hopping $t_0$ and are therefore independent of the
absolute energy scale; the connection to specific molecular platforms and
experimental energy scales is discussed in Sec.~II.E.
Two experimental signatures distinguish the chiral molecular realization
from a generic SSH chain: the chirality-renormalized enhancement of the
bulk topological gap by the Dzyaloshinskii--Moriya interaction, and the
opposite molecular handedness of the two boundary modes. These provide
concrete targets for future experiments as the laser cooling and
optical tweezer trapping of chiral polyatomic species continues to
advance.

In summary, dimerized chiral molecular arrays support SSH-like
topological phases with nontrivial winding, bulk-gap closing at the
transition point, and boundary-localized edge modes in the topological
regime. The single-chain and ladder constructions studied here establish
this platform as a promising candidate for controllable topological
phases and chirality-addressable edge states in interacting molecular
quantum matter.

\section*{ACKNOWLEDGMENTS}
The authors would like to acknowledge the financial support from the Quantum Science Center, a National Quantum Information Science Research Center of the U.S. Department of Energy (DOE), operated at Oak Ridge National Laboratory (ORNL).

\bibliography{apssamp}
\appendix
\onecolumngrid
\renewcommand{\thesection}{\Alph{section}}
\renewcommand{\theequation}{\thesection\arabic{equation}}
\section{Gauge Transformation, Jordan--Wigner Mapping, and the Interacting SSH Chain}
\label{app:gauge_jw}

This appendix derives the interacting SSH Hamiltonian used in Sec.~\ref{sec:model}. We start from the chiral XXZ spin Hamiltonian, rewrite the transverse part in ladder-operator form, remove the Dzyaloshinskii--Moriya (DM) phase by a site-dependent gauge transformation, and then apply the Jordan--Wigner transformation. Finally, we introduce the dimerized molecular spacing that produces the two SSH hopping amplitudes \(t_1\) and \(t_2\).

\subsection{Starting spin Hamiltonian}

In the Pauli-matrix convention, the effective open-chain spin Hamiltonian is
\begin{equation}
\hat H_{\mathrm{spin}}
=
\sum_{j=1}^{N-1}
\left[
J_{xy,j}
(\hat\sigma_j^x\hat\sigma_{j+1}^x+\hat\sigma_j^y\hat\sigma_{j+1}^y)
-
D_j
(\hat\sigma_j^x\hat\sigma_{j+1}^y-\hat\sigma_j^y\hat\sigma_{j+1}^x)
+
J_{z,j}\hat\sigma_j^z\hat\sigma_{j+1}^z
\right]
+
h\sum_{j=1}^{N}\hat\sigma_j^z .
\label{eqA:Hspin}
\end{equation}
Here \(\hat\sigma_j^\alpha\) are Pauli matrices acting on site \(j\). The coefficient \(J_{xy,j}\) is the symmetric transverse exchange, \(D_j\) is the DM coupling, \(J_{z,j}\) is the Ising interaction, and \(h\equiv h(\bar r)\) is the effective longitudinal field evaluated at the mean intermolecular separation.

The microscopic couplings are
\begin{align}
J_{xy,j}
&=
-\frac{\Omega(r_j)}{2}\Re(C_{d_1}),
\\
D_j
&=
+\frac{\Omega(r_j)}{2}\Im(C_{d_1}),
\\
J_{z,j}
&=
\frac{\Omega(r_j)}{4}
\big[(C_2+C_3)-(C_1+C_4)\big],
\\
h(\bar r)
&=
\frac{
2(E_\uparrow-E_\downarrow)
+
\Omega(\bar r)(C_1-C_4)
}{4}.
\end{align}
The dipolar scale is
\begin{equation}
\Omega(r_j)
=
\frac{1}{h_{\mathrm P}}
\frac{|\mathbf d|^2}{4\pi\epsilon_0 r_j^3},
\qquad
\Omega(r_j)\propto r_j^{-3},
\end{equation}
where \(h_{\mathrm P}\) is Planck's constant, \(\mathbf d\) is the molecular dipole moment, and \(\epsilon_0\) is the vacuum permittivity. The coefficients \(C_{d_1}\) and \(C_i\) are dressed-state dipole matrix elements, while \(E_\uparrow\) and \(E_\downarrow\) are dressed-state energies.

A key point is that \(J_{xy,j}\) and \(D_j\) contain the same distance-dependent factor \(\Omega(r_j)\). Therefore,
\begin{equation}
\frac{D_j}{J_{xy,j}}
=
-
\frac{\Im(C_{d_1})}{\Re(C_{d_1})},
\label{eqA:D_over_J}
\end{equation}
which is independent of the bond length \(r_j\). This allows the DM phase to be removed from every nearest-neighbor bond of an open chain by a site-dependent rotation about the \(z\)-axis.

\subsection{Transverse exchange in ladder-operator form}

We now isolate the transverse part of Eq.~\eqref{eqA:Hspin}. For a single bond \((j,j+1)\), define
\begin{align}
\hat H_{j}^{xy+DM}
&=
J_{xy,j}
(\hat\sigma_j^x\hat\sigma_{j+1}^x+\hat\sigma_j^y\hat\sigma_{j+1}^y)
\nonumber\\
&\quad
-
D_j
(\hat\sigma_j^x\hat\sigma_{j+1}^y-\hat\sigma_j^y\hat\sigma_{j+1}^x).
\end{align}
This is the part of the Hamiltonian containing the symmetric transverse exchange and the antisymmetric DM interaction on bond \(j\).

Introduce the ladder operators
\[
\hat\sigma_j^\pm
=
\frac{1}{2}
(\hat\sigma_j^x\pm i\hat\sigma_j^y),
\]
so that
\[
\hat\sigma_j^x=\hat\sigma_j^++\hat\sigma_j^-,
\qquad
\hat\sigma_j^y=-i(\hat\sigma_j^+-\hat\sigma_j^-).
\]
Using these definitions, the symmetric transverse exchange becomes
\begin{align}
\hat\sigma_j^x\hat\sigma_{j+1}^x
+
\hat\sigma_j^y\hat\sigma_{j+1}^y
&=
2
\left(
\hat\sigma_j^+\hat\sigma_{j+1}^-
+
\hat\sigma_j^-\hat\sigma_{j+1}^+
\right),
\end{align}
while the DM combination becomes
\begin{align}
\hat\sigma_j^x\hat\sigma_{j+1}^y
-
\hat\sigma_j^y\hat\sigma_{j+1}^x
&=
2i
\left(
\hat\sigma_j^+\hat\sigma_{j+1}^-
-
\hat\sigma_j^-\hat\sigma_{j+1}^+
\right).
\end{align}
Substituting these identities into \(\hat H_j^{xy+DM}\), we obtain
\begin{align}
\hat H_j^{xy+DM}
&=
2(J_{xy,j}-iD_j)
\hat\sigma_j^+\hat\sigma_{j+1}^-
+
2(J_{xy,j}+iD_j)
\hat\sigma_j^-\hat\sigma_{j+1}^+ .
\label{eqA:HxyDM_ladder}
\end{align}
Thus the transverse exchange and the DM interaction combine into complex hopping-like spin-flip amplitudes.

It is useful to write the complex coupling in polar form:
\begin{equation}
J_{xy,j}+iD_j
=
\widetilde J_{xy,j}e^{i\theta_j},
\qquad
J_{xy,j}-iD_j
=
\widetilde J_{xy,j}e^{-i\theta_j},
\end{equation}
where
\begin{equation}
\widetilde J_{xy,j}
=
\sqrt{J_{xy,j}^2+D_j^2},
\qquad
\theta_j
=
\arg(J_{xy,j}+iD_j).
\label{eqA:Jtilde_theta}
\end{equation}
The magnitude \(\widetilde J_{xy,j}\) is the chirality-renormalized transverse exchange. Because \(D_j/J_{xy,j}\) is independent of \(r_j\), the phase \(\theta_j\) is bond-independent. We therefore write simply
\[
\theta_j=\theta .
\]

\subsection{Gauge transformation}

The complex phase in Eq.~\eqref{eqA:HxyDM_ladder} can be removed by a site-dependent rotation about the \(z\)-axis. We define
\begin{equation}
U
=
\prod_{j=1}^{N}
\exp\!\left[
-\frac{i}{2}\phi_j\hat\sigma_j^z
\right].
\label{eqA:U}
\end{equation}
The factor \(1/2\) appears because the generator of a spin rotation is \(\hat\sigma_j^z/2\). Using
\[
[\hat\sigma_j^z,\hat\sigma_j^+]=2\hat\sigma_j^+,
\qquad
[\hat\sigma_j^z,\hat\sigma_j^-]=-2\hat\sigma_j^-,
\]
one obtains
\[
U\hat\sigma_j^+U^\dagger=e^{-i\phi_j}\hat\sigma_j^+,
\qquad
U\hat\sigma_j^-U^\dagger=e^{+i\phi_j}\hat\sigma_j^-,
\qquad
U\hat\sigma_j^zU^\dagger=\hat\sigma_j^z .
\]
Therefore,
\begin{align}
U
(\hat\sigma_j^+\hat\sigma_{j+1}^-)
U^\dagger
&=
e^{i(\phi_{j+1}-\phi_j)}
\hat\sigma_j^+\hat\sigma_{j+1}^-,
\\
U
(\hat\sigma_j^-\hat\sigma_{j+1}^+)
U^\dagger
&=
e^{-i(\phi_{j+1}-\phi_j)}
\hat\sigma_j^-\hat\sigma_{j+1}^+ .
\end{align}
Applying this to Eq.~\eqref{eqA:HxyDM_ladder} gives
\begin{align}
U\hat H_j^{xy+DM}U^\dagger
&=
2\widetilde J_{xy,j}
e^{i(\phi_{j+1}-\phi_j-\theta)}
\hat\sigma_j^+\hat\sigma_{j+1}^-
\nonumber\\
&\quad
+
2\widetilde J_{xy,j}
e^{-i(\phi_{j+1}-\phi_j-\theta)}
\hat\sigma_j^-\hat\sigma_{j+1}^+ .
\end{align}
The phase disappears if
\begin{equation}
\phi_{j+1}-\phi_j=\theta .
\label{eqA:gauge_condition}
\end{equation}
Since \(\theta\) is bond-independent, Eq.~\eqref{eqA:gauge_condition} can be satisfied on every open-chain bond by choosing
\[
\phi_j=j\theta .
\]
With this choice,
\begin{align}
U\hat H_j^{xy+DM}U^\dagger
&=
2\widetilde J_{xy,j}
\left(
\hat\sigma_j^+\hat\sigma_{j+1}^-
+
\hat\sigma_j^-\hat\sigma_{j+1}^+
\right)
\nonumber\\
&=
\widetilde J_{xy,j}
(\hat\sigma_j^x\hat\sigma_{j+1}^x+\hat\sigma_j^y\hat\sigma_{j+1}^y).
\end{align}
The Ising and field terms are unchanged because they contain only \(\hat\sigma^z\). Hence the gauge-rotated spin Hamiltonian is
\begin{align}
\hat H_{\mathrm{rot}}
&=
\sum_{j=1}^{N-1}
\left[
\widetilde J_{xy,j}
(\hat\sigma_j^x\hat\sigma_{j+1}^x+\hat\sigma_j^y\hat\sigma_{j+1}^y)
+
J_{z,j}\hat\sigma_j^z\hat\sigma_{j+1}^z
\right]
+
h\sum_{j=1}^{N}\hat\sigma_j^z .
\label{eqA:Hrot}
\end{align}
The DM interaction has therefore been absorbed into the magnitude \(\widetilde J_{xy,j}\). For a periodic chain, the total accumulated phase around the ring appears as a boundary twist. In the thermodynamic limit this twist only shifts the allowed momenta and does not change the bulk gap, the SSH winding criterion, or the open-boundary edge-state analysis used here.

\subsection{Jordan--Wigner transformation}

We now map Eq.~\eqref{eqA:Hrot} to spinless fermions. The Jordan--Wigner transformation is
\begin{align}
\hat\sigma_j^+
&=
\hat c_j^\dagger
\prod_{\ell<j}
(1-2\hat n_\ell),
\\
\hat\sigma_j^-
&=
\prod_{\ell<j}
(1-2\hat n_\ell)
\hat c_j,
\\
\hat\sigma_j^z
&=
2\hat n_j-1,
\qquad
\hat n_j=\hat c_j^\dagger\hat c_j .
\end{align}
The string operators ensure that fermions on different sites anticommute. For nearest-neighbor terms on an open chain, the strings cancel. Explicitly,
\begin{align}
\hat\sigma_j^+\hat\sigma_{j+1}^-
&=
\hat c_j^\dagger
\prod_{\ell<j}(1-2\hat n_\ell)
\prod_{\ell'<j+1}(1-2\hat n_{\ell'})
\hat c_{j+1}
\nonumber\\
&=
\hat c_j^\dagger
(1-2\hat n_j)
\hat c_{j+1}.
\end{align}
Since
\[
\hat c_j^\dagger\hat n_j
=
\hat c_j^\dagger \hat c_j^\dagger \hat c_j
=
0,
\]
we have
\[
\hat c_j^\dagger(1-2\hat n_j)=\hat c_j^\dagger,
\]
and therefore
\begin{equation}
\hat\sigma_j^+\hat\sigma_{j+1}^-
=
\hat c_j^\dagger\hat c_{j+1}.
\end{equation}
Similarly,
\begin{equation}
\hat\sigma_j^-\hat\sigma_{j+1}^+
=
\hat c_{j+1}^\dagger\hat c_j .
\end{equation}
Thus the transverse part becomes
\begin{align}
\widetilde J_{xy,j}
(\hat\sigma_j^x\hat\sigma_{j+1}^x+\hat\sigma_j^y\hat\sigma_{j+1}^y)
&=
2\widetilde J_{xy,j}
(
\hat\sigma_j^+\hat\sigma_{j+1}^-
+
\hat\sigma_j^-\hat\sigma_{j+1}^+
)
\nonumber\\
&\longrightarrow
2\widetilde J_{xy,j}
(
\hat c_j^\dagger\hat c_{j+1}
+
\hat c_{j+1}^\dagger\hat c_j
).
\end{align}
Therefore the fermionic hopping amplitude is
\begin{equation}
t_j=2\widetilde J_{xy,j}
=
2\sqrt{J_{xy,j}^2+D_j^2}.
\label{eqA:tj_def}
\end{equation}
The factor of \(2\) comes from using Pauli matrices in the spin Hamiltonian.

The Ising term maps as
\begin{align}
J_{z,j}\hat\sigma_j^z\hat\sigma_{j+1}^z
&=
J_{z,j}(2\hat n_j-1)(2\hat n_{j+1}-1)
\nonumber\\
&=
4J_{z,j}\hat n_j\hat n_{j+1}
-
2J_{z,j}\hat n_j
-
2J_{z,j}\hat n_{j+1}
+
J_{z,j}.
\end{align}
The field term gives
\begin{align}
h\sum_{j=1}^{N}\hat\sigma_j^z
&=
h\sum_{j=1}^{N}(2\hat n_j-1)
\nonumber\\
&=
2h\sum_{j=1}^{N}\hat n_j
-
hN .
\end{align}
Combining all pieces, the exact open-chain fermionic Hamiltonian is
\begin{align}
\hat H_{\mathrm{ferm}}^{\mathrm{OBC}}
&=
\sum_{j=1}^{N-1}
t_j
(\hat c_j^\dagger\hat c_{j+1}
+
\hat c_{j+1}^\dagger\hat c_j)
+
4\sum_{j=1}^{N-1}
J_{z,j}\hat n_j\hat n_{j+1}
\nonumber\\
&\quad
-
2\sum_{j=1}^{N-1}
J_{z,j}
(\hat n_j+\hat n_{j+1})
+
2h\sum_{j=1}^{N}\hat n_j
+
\sum_{j=1}^{N-1}J_{z,j}
-
hN .
\label{eqA:Hfermion_exact}
\end{align}

For the reduced bulk theory used in the main text, we keep the dimerization explicitly in the transverse hopping amplitudes and take the longitudinal interaction at the mean separation,
\[
J_z\equiv J_z(\bar r).
\]
With this replacement, the linear density part generated by the Ising term is
\[
-2J_z\sum_{j=1}^{N-1}(\hat n_j+\hat n_{j+1}).
\]
The sum can be written exactly as
\[
\sum_{j=1}^{N-1}(\hat n_j+\hat n_{j+1})
=
2\sum_{j=1}^{N}\hat n_j-\hat n_1-\hat n_N .
\]
Therefore,
\begin{align}
-2J_z\sum_{j=1}^{N-1}(\hat n_j+\hat n_{j+1})
&=
-4J_z\sum_{j=1}^{N}\hat n_j
+
2J_z(\hat n_1+\hat n_N).
\end{align}
The second term is a boundary density correction. It appears because the two end sites belong to only one bond, while interior sites belong to two bonds. In the bulk-effective Hamiltonian this boundary correction is omitted because it is a local end potential and does not change the thermodynamic bulk invariant. Its possible effect on finite-size edge energies is separated from the bulk topological classification.

The bulk-effective fermionic Hamiltonian is therefore
\begin{align}
\hat H_{\mathrm{bulk}}
&=
\sum_{j=1}^{N-1}
t_j
(\hat c_j^\dagger\hat c_{j+1}+\mathrm{h.c.})
+
4J_z
\sum_{j=1}^{N-1}
\hat n_j\hat n_{j+1}
\nonumber\\
&\quad
+
\mu_{\mathrm{bulk}}
\sum_{j=1}^{N}
\hat n_j
+
E_{\mathrm{const}},
\label{eqA:Hbulk}
\end{align}
where
\begin{equation}
\mu_{\mathrm{bulk}}=2h-4J_z,
\qquad
E_{\mathrm{const}}=J_z(N-1)-hN .
\label{eqA:mu_Econst}
\end{equation}
The constant \(E_{\mathrm{const}}\) comes from the \(+J_z\) contribution on each Ising bond and the \(-h\) contribution from the field. It shifts all energies uniformly and has no effect on eigenstates, gaps, winding numbers, or edge localization.

The interaction term
\[
4J_z\hat n_j\hat n_{j+1}
=
4J_z
\hat c_j^\dagger\hat c_j
\hat c_{j+1}^\dagger\hat c_{j+1}
\]
contains four fermionic operators. This is the many-body term that is decoupled in Appendix~\ref{app:mf_decoupling}.

\subsection{Dimerization and interacting SSH form}

The SSH structure is introduced by alternating the molecular spacing along the chain. We use the convention
\begin{equation}
r_j=\bar r-(-1)^j\delta r .
\label{eqA:rj_convention}
\end{equation}
Thus
\[
j\ \mathrm{odd}:\quad r_j=\bar r+\delta r,
\qquad
j\ \mathrm{even}:\quad r_j=\bar r-\delta r.
\]
Since \(\Omega(r)\propto r^{-3}\), the shorter bond has the larger hopping. Using Eq.~\eqref{eqA:tj_def}, the two alternating hoppings are
\begin{align}
t_1
&=
2\widetilde J_{xy}(\bar r)
\left(
\frac{\bar r}{\bar r+\delta r}
\right)^3,
\nonumber\\
t_2
&=
2\widetilde J_{xy}(\bar r)
\left(
\frac{\bar r}{\bar r-\delta r}
\right)^3.
\label{eqA:t1t2}
\end{align}
We identify odd bonds as intracell bonds and even bonds as intercell bonds:
\[
t_j=
\begin{cases}
t_1, & j\ \mathrm{odd},\\[3pt]
t_2, & j\ \mathrm{even}.
\end{cases}
\]
With this convention, positive \(\delta r\) gives \(t_2>t_1\), which is the topological SSH pattern for the termination used in this work.

Substituting these alternating hoppings into Eq.~\eqref{eqA:Hbulk}, we obtain the interacting SSH Hamiltonian
\begin{align}
\hat H_{\mathrm{SSH}}^{\mathrm{int}}
&=
\sum_{j\ \mathrm{odd}}
t_1
(\hat c_j^\dagger\hat c_{j+1}+\mathrm{h.c.})
+
\sum_{j\ \mathrm{even}}
t_2
(\hat c_j^\dagger\hat c_{j+1}+\mathrm{h.c.})
\nonumber\\
&\quad
+
4J_z
\sum_{j=1}^{N-1}
\hat n_j\hat n_{j+1}
+
\mu_{\mathrm{bulk}}
\sum_{j=1}^{N}
\hat n_j
+
E_{\mathrm{const}} .
\label{eqA:HSSH_int}
\end{align}
This is the interacting fermionic SSH model used as the starting point for the mean-field treatment.

Finally, half filling corresponds to the zero-magnetization sector of the spin chain. Since
\[
\hat\sigma_j^z=2\hat n_j-1,
\]
we have
\[
\sum_{j=1}^{N}\langle \hat\sigma_j^z\rangle
=
2\sum_{j=1}^{N}\langle \hat n_j\rangle-N.
\]
Thus zero total magnetization implies
\[
\sum_{j=1}^{N}\langle \hat n_j\rangle=\frac{N}{2}.
\]
Equivalently, the fermionic filling is
\[
N_f=\frac{N}{2}.
\]
This is the half-filled sector used in the Hartree--Fock decoupling and in the subsequent topological analysis.
\section{Hartree--Fock Mean-Field Decoupling and Renormalized SSH Parameters}
\label{app:mf_decoupling}

This appendix derives the quadratic mean-field Hamiltonian used in Sec.~\ref{sec:model}. The starting point is the interacting SSH Hamiltonian obtained in Appendix~\ref{app:gauge_jw}. At this stage the hopping part already has the dimerized SSH form, while the density--density term inherited from the Ising interaction is quartic in fermionic operators. The Hartree--Fock approximation replaces this quartic term by a quadratic operator whose coefficients are determined self-consistently.

The bulk-effective interacting fermionic Hamiltonian is
\begin{align}
\hat H_{\mathrm{bulk}}
&=
\sum_{j=1}^{N-1}
t_j
\left(
\hat c_j^\dagger\hat c_{j+1}
+
\hat c_{j+1}^\dagger\hat c_j
\right)
+
4J_z
\sum_{j=1}^{N-1}
\hat n_j\hat n_{j+1}
+
\mu_{\mathrm{bulk}}
\sum_{j=1}^{N}
\hat n_j
+
E_{\mathrm{const}},
\label{eqB:Hbulk}
\end{align}
where
\[
\hat n_j=\hat c_j^\dagger\hat c_j,
\qquad
\mu_{\mathrm{bulk}}=2h-4J_z,
\qquad
E_{\mathrm{const}}=J_z(N-1)-hN.
\]
The hopping alternates as
\[
t_j=
\begin{cases}
t_1, & j\ \mathrm{odd},\\[3pt]
t_2, & j\ \mathrm{even},
\end{cases}
\]
with \(t_1\) the intracell hopping and \(t_2\) the intercell hopping. The constant \(E_{\mathrm{const}}\) is the Jordan--Wigner constant derived in Appendix~\ref{app:gauge_jw}. It is kept for energy accounting, but it does not affect the single-particle eigenvectors, gaps, winding number, or edge localization.

The only non-quadratic term in Eq.~\eqref{eqB:Hbulk} is
\[
4J_z\hat n_j\hat n_{j+1}
=
4J_z
\hat c_j^\dagger\hat c_j
\hat c_{j+1}^\dagger\hat c_{j+1}.
\]
This contains four fermionic operators and therefore cannot be represented as a single-particle matrix before approximation. The mean-field treatment replaces this quartic interaction by density and exchange channels.

We work at half filling. In the spin language this corresponds to the zero-magnetization sector, because
\[
\hat\sigma_j^z=2\hat n_j-1
\]
implies
\[
\sum_{j=1}^{N}\langle\hat\sigma_j^z\rangle
=
2\sum_{j=1}^{N}\langle\hat n_j\rangle-N.
\]
Thus zero total magnetization gives
\[
\sum_{j=1}^{N}\langle\hat n_j\rangle
=
\frac{N}{2}.
\]
In the translationally invariant bulk mean-field treatment, this is implemented as
\[
\langle\hat n_j\rangle=\frac12.
\]

Because the SSH chain has two inequivalent bond types, the bond coherence must also be allowed to depend on the bond parity. We define
\begin{equation}
\chi_1
=
\frac{1}{\mathcal N_1}
\sum_{j\ \mathrm{odd}}
\left\langle
\hat c_j^\dagger\hat c_{j+1}
\right\rangle,
\qquad
\chi_2
=
\frac{1}{\mathcal N_2}
\sum_{j\ \mathrm{even}}
\left\langle
\hat c_j^\dagger\hat c_{j+1}
\right\rangle .
\label{eqB:chi_def}
\end{equation}
Here \(\mathcal N_1\) and \(\mathcal N_2\) are the numbers of odd and even bonds included in the corresponding averages. For an open chain with \(N=2N_c\), if the first bond is odd, \(\mathcal N_1=N_c\) and \(\mathcal N_2=N_c-1\). This difference is a boundary effect and disappears in the thermodynamic limit. The quantities \(\chi_1\) and \(\chi_2\) are the quantum amplitudes for hopping across the two SSH bond types. In the spin language, they correspond to nearest-neighbor flip-flop correlations in the gauge-rotated frame.

We now decouple the interaction on a single bond. The exact interaction is
\[
\hat n_j\hat n_{j+1}
=
\hat c_j^\dagger\hat c_j
\hat c_{j+1}^\dagger\hat c_{j+1}.
\]
The Hartree--Fock approximation keeps the density contractions and exchange contractions that leave a quadratic operator, and subtracts the corresponding constant pieces to avoid double counting. The density contractions are
\[
\langle \hat c_j^\dagger\hat c_j\rangle
=
\langle \hat n_j\rangle,
\qquad
\langle \hat c_{j+1}^\dagger\hat c_{j+1}\rangle
=
\langle \hat n_{j+1}\rangle .
\]
They give the Hartree contribution
\[
\langle\hat n_j\rangle \hat n_{j+1}
+
\hat n_j\langle\hat n_{j+1}\rangle
-
\langle\hat n_j\rangle\langle\hat n_{j+1}\rangle .
\]
The exchange contractions are
\[
\left\langle
\hat c_j^\dagger\hat c_{j+1}
\right\rangle,
\qquad
\left\langle
\hat c_{j+1}^\dagger\hat c_j
\right\rangle .
\]
They give the Fock contribution
\[
-
\left\langle
\hat c_j^\dagger\hat c_{j+1}
\right\rangle
\hat c_{j+1}^\dagger\hat c_j
-
\left\langle
\hat c_{j+1}^\dagger\hat c_j
\right\rangle
\hat c_j^\dagger\hat c_{j+1}
+
\left\langle
\hat c_j^\dagger\hat c_{j+1}
\right\rangle
\left\langle
\hat c_{j+1}^\dagger\hat c_j
\right\rangle .
\]
The minus signs in the Fock terms are the fermionic exchange signs. They appear because the exchanged operators must be anticommuted when forming the intersite contractions.

Combining the density and exchange channels gives
\begin{align}
\hat n_j\hat n_{j+1}
&\approx
\langle\hat n_j\rangle\hat n_{j+1}
+
\hat n_j\langle\hat n_{j+1}\rangle
-
\langle\hat n_j\rangle\langle\hat n_{j+1}\rangle
\nonumber\\
&\quad
-
\left\langle
\hat c_j^\dagger\hat c_{j+1}
\right\rangle
\hat c_{j+1}^\dagger\hat c_j
-
\left\langle
\hat c_{j+1}^\dagger\hat c_j
\right\rangle
\hat c_j^\dagger\hat c_{j+1}
\nonumber\\
&\quad
+
\left\langle
\hat c_j^\dagger\hat c_{j+1}
\right\rangle
\left\langle
\hat c_{j+1}^\dagger\hat c_j
\right\rangle .
\label{eqB:HF_general}
\end{align}

At half filling,
\[
\langle\hat n_j\rangle
=
\langle\hat n_{j+1}\rangle
=
\frac12.
\]
After the gauge transformation of Appendix~\ref{app:gauge_jw}, the effective hopping amplitudes are real. We therefore use a real bond-order ansatz,
\[
\chi_j
=
\left\langle
\hat c_j^\dagger\hat c_{j+1}
\right\rangle
=
\left\langle
\hat c_{j+1}^\dagger\hat c_j
\right\rangle
\in\mathbb R,
\]
with
\[
\chi_j=
\begin{cases}
\chi_1, & j\ \mathrm{odd},\\[3pt]
\chi_2, & j\ \mathrm{even}.
\end{cases}
\]
Substituting these values into Eq.~\eqref{eqB:HF_general} gives
\[
\hat n_j\hat n_{j+1}
\approx
\frac12\hat n_{j+1}
+
\frac12\hat n_j
-
\frac14
-
\chi_j\hat c_{j+1}^\dagger\hat c_j
-
\chi_j\hat c_j^\dagger\hat c_{j+1}
+
\chi_j^2 .
\]
Multiplying by \(4J_z\), the interaction on one bond becomes
\begin{equation}
4J_z\hat n_j\hat n_{j+1}
\approx
2J_z(\hat n_j+\hat n_{j+1})
-
4J_z\chi_j
(\hat c_j^\dagger\hat c_{j+1}+\hat c_{j+1}^\dagger\hat c_j)
-
J_z
+
4J_z\chi_j^2 .
\label{eqB:single_bond_decoupling}
\end{equation}
This is the central mean-field replacement. The first term is the Hartree density shift, the second term is the Fock correction to the hopping, and the last two terms are constants.

We now collect these contributions over the whole open chain. First consider the Hartree part:
\[
\hat H_{\mathrm{H}}
=
2J_z
\sum_{j=1}^{N-1}
(\hat n_j+\hat n_{j+1}).
\]
The sum can be written exactly as
\[
\sum_{j=1}^{N-1}
(\hat n_j+\hat n_{j+1})
=
\hat n_1+2\hat n_2+\cdots+2\hat n_{N-1}+\hat n_N
=
2\sum_{j=1}^{N}\hat n_j-\hat n_1-\hat n_N .
\]
Hence
\[
\hat H_{\mathrm{H}}
=
4J_z
\sum_{j=1}^{N}\hat n_j
-
2J_z(\hat n_1+\hat n_N).
\]
The second term is a boundary correction. In the bulk mean-field Hamiltonian used for band structure and topology, this boundary contribution is omitted. This boundary Hartree term should be distinguished from the exact
Jordan--Wigner boundary density correction derived in Appendix~A.
If the exact open-chain boundary term \(+2J_z(\hat n_1+\hat n_N)\)
is retained, the half-filled Hartree boundary contribution
\(-2J_z(\hat n_1+\hat n_N)\) cancels it. In the bulk-effective
mean-field theory used here, both local end contributions are omitted
so that the translationally invariant bulk Hamiltonian determines the
band structure and topological invariant. The Hartree term is taken as
\[
\hat H_{\mathrm{H}}^{\mathrm{bulk}}
=
4J_z
\sum_{j=1}^{N}\hat n_j .
\]

Adding this to the original density term in Eq.~\eqref{eqB:Hbulk} gives
\begin{equation}
\mu_{\mathrm{eff}}
=
\mu_{\mathrm{bulk}}+4J_z
=
(2h-4J_z)+4J_z
=
2h .
\label{eqB:mu_eff}
\end{equation}
Thus the bulk linear density term generated by the Jordan--Wigner mapping of the Ising interaction is canceled by the Hartree contribution of the same interaction. The remaining \(\mu_{\mathrm{eff}}\) is a uniform spectral shift.

Next consider the Fock part. On bond \(j\), the original hopping and the Fock correction have the same operator structure:
\[
t_j
(\hat c_j^\dagger\hat c_{j+1}+\mathrm{h.c.})
-
4J_z\chi_j
(\hat c_j^\dagger\hat c_{j+1}+\mathrm{h.c.})
=
(t_j-4J_z\chi_j)
(\hat c_j^\dagger\hat c_{j+1}+\mathrm{h.c.}).
\]
Therefore,
\[
t_j^{\mathrm{eff}}=t_j-4J_z\chi_j .
\]
For the two SSH bond types,
\begin{equation}
t_1^{\mathrm{eff}}
=
t_1-4J_z\chi_1,
\qquad
t_2^{\mathrm{eff}}
=
t_2-4J_z\chi_2 .
\label{eqB:teff}
\end{equation}
These are the renormalized SSH hopping amplitudes used throughout the bulk and open-boundary calculations.

The constants should be separated into two parts. The first is the Jordan--Wigner constant already present in Eq.~\eqref{eqB:Hbulk},
\[
E_{\mathrm{const}}=J_z(N-1)-hN.
\]
The second is generated by the Hartree--Fock decoupling of the interaction:
\[
E_{\mathrm{HF}}^{\mathrm{const}}
=
\sum_{j=1}^{N-1}
(-J_z+4J_z\chi_j^2).
\]
Using the two bond types,
\begin{equation}
E_{\mathrm{HF}}^{\mathrm{const}}
=
-J_z(N-1)
+
4J_z
\left(
\mathcal N_1\chi_1^2+\mathcal N_2\chi_2^2
\right).
\label{eqB:EHF_const}
\end{equation}
The total constant is therefore
\[
E_{\mathrm{tot}}^{\mathrm{const}}
=
E_{\mathrm{const}}+E_{\mathrm{HF}}^{\mathrm{const}}.
\]
These constants are relevant for comparing total mean-field energies, but they do not enter the single-particle eigenvectors or the topological invariant.

Collecting the hopping, density, and constant terms, the full mean-field Hamiltonian becomes
\begin{equation}
\hat H_{\mathrm{MF}}
=
\sum_{j\ \mathrm{odd}}
t_1^{\mathrm{eff}}
(\hat c_j^\dagger\hat c_{j+1}+\mathrm{h.c.})
+
\sum_{j\ \mathrm{even}}
t_2^{\mathrm{eff}}
(\hat c_j^\dagger\hat c_{j+1}+\mathrm{h.c.})
+
\mu_{\mathrm{eff}}
\sum_{j=1}^{N}\hat n_j
+
E_{\mathrm{tot}}^{\mathrm{const}} .
\label{eqB:HMF_full}
\end{equation}
Since \(\mu_{\mathrm{eff}}\) is proportional to the identity in the single-particle Hamiltonian and \(E_{\mathrm{tot}}^{\mathrm{const}}\) is an overall energy offset, the reduced Hamiltonian controlling topology and edge localization is
\begin{equation}
\hat H_{\mathrm{MF}}'
=
\sum_{j\ \mathrm{odd}}
t_1^{\mathrm{eff}}
(\hat c_j^\dagger\hat c_{j+1}+\mathrm{h.c.})
+
\sum_{j\ \mathrm{even}}
t_2^{\mathrm{eff}}
(\hat c_j^\dagger\hat c_{j+1}+\mathrm{h.c.}) .
\label{eqB:HMF_reduced}
\end{equation}
The interacting problem has therefore been reduced to a quadratic SSH Hamiltonian, but with hopping amplitudes determined by the interacting ground state.

For open boundary conditions, Eq.~\eqref{eqB:HMF_reduced} is represented by an \(N\times N\) tridiagonal matrix with
\[
(H_{\mathrm{MF}}')_{j,j+1}
=
(H_{\mathrm{MF}}')_{j+1,j}
=
\begin{cases}
t_1^{\mathrm{eff}}, & j\ \mathrm{odd},\\[3pt]
t_2^{\mathrm{eff}}, & j\ \mathrm{even}.
\end{cases}
\]
Let \(\psi_m(j)\) be the amplitude of the \(m\)-th single-particle eigenstate on site \(j\),
\[
H_{\mathrm{MF}}'\psi_m=E_m\psi_m.
\]
At half filling, the lowest \(N/2\) single-particle states are occupied. The bond orders are recomputed from these occupied eigenstates:
\begin{align}
\chi_1
&=
\frac{1}{\mathcal N_1}
\sum_{m\in\mathrm{occ}}
\sum_{j\ \mathrm{odd}}
\psi_m^*(j)\psi_m(j+1),
\nonumber\\
\chi_2
&=
\frac{1}{\mathcal N_2}
\sum_{m\in\mathrm{occ}}
\sum_{j\ \mathrm{even}}
\psi_m^*(j)\psi_m(j+1).
\label{eqB:SC_OBC}
\end{align}
These equations close the Hartree--Fock problem. In practice, one starts from trial values of \(\chi_1\) and \(\chi_2\), constructs \(t_1^{\mathrm{eff}}\) and \(t_2^{\mathrm{eff}}\), diagonalizes \(H_{\mathrm{MF}}'\), fills the lowest \(N/2\) states, recomputes \(\chi_1\) and \(\chi_2\), and repeats until convergence.

At convergence, \(\chi_1^\star\) and \(\chi_2^\star\) are the self-consistent Hartree--Fock bond orders. The interaction then enters the subsequent topological analysis only through \(t_1^{\mathrm{eff}}\) and \(t_2^{\mathrm{eff}}\). These effective hopping amplitudes are used in Appendix~\ref{app:topology_edge} to compute the bulk winding number, the bulk gap, and the open-boundary edge states.
\section{Bulk Topology, Periodic-Boundary Self-Consistency, and Open-Boundary Edge States}
\label{app:topology_edge}

This appendix derives the bulk topology and open-boundary edge-state properties of the self-consistent mean-field SSH Hamiltonian obtained in Appendix~\ref{app:mf_decoupling}. The uniform density shift \(\mu_{\mathrm{eff}}=2h\) is first subtracted, because it is proportional to the identity in the single-particle Hamiltonian and therefore does not affect eigenvectors, winding numbers, localization lengths, or finite-size edge-state splittings. When absolute energies are required, this shift is restored by adding \(2h\) to every single-particle eigenvalue.

\subsection{Periodic-boundary formulation}

After Hartree--Fock decoupling, the reduced mean-field Hamiltonian is
\begin{equation}
\hat H_{\mathrm{MF}}'
=
\sum_{j\ \mathrm{odd}}t_1^{\mathrm{eff}}(\hat c_j^\dagger\hat c_{j+1}+\mathrm{h.c.})
+
\sum_{j\ \mathrm{even}}t_2^{\mathrm{eff}}(\hat c_j^\dagger\hat c_{j+1}+\mathrm{h.c.}),
\label{eqC:HMF_reduced}
\end{equation}
where
\[
t_1^{\mathrm{eff}}=t_1-4J_z\chi_1,
\qquad
t_2^{\mathrm{eff}}=t_2-4J_z\chi_2 .
\]
We take \(N=2N_c\), where \(N_c\) is the number of two-site unit cells. The odd bonds are the intracell bonds and the even bonds are the intercell bonds. With the dimerization convention used in Appendix~\ref{app:gauge_jw}, the topological termination corresponds to
\[
|t_2^{\mathrm{eff}}|>|t_1^{\mathrm{eff}}|.
\]

Under periodic boundary conditions, the chain is closed by identifying
\[
\hat c_{2N_c+1}\equiv \hat c_1.
\]
It is convenient to label the two sites in unit cell \(m\) as the odd site \(2m-1\) and the even site \(2m\). Then Eq.~\eqref{eqC:HMF_reduced} becomes
\begin{align}
\hat H_{\mathrm{MF}}^{\mathrm{PBC}}
&=
\sum_{m=1}^{N_c}
t_1^{\mathrm{eff}}
\left(
\hat c_{2m-1}^\dagger\hat c_{2m}
+
\hat c_{2m}^\dagger\hat c_{2m-1}
\right)
\nonumber\\
&\quad+
\sum_{m=1}^{N_c}
t_2^{\mathrm{eff}}
\left(
\hat c_{2m}^\dagger\hat c_{2m+1}
+
\hat c_{2m+1}^\dagger\hat c_{2m}
\right).
\label{eqC:HPBC_real}
\end{align}
The first sum describes hopping inside each unit cell. The second sum describes hopping from the even site of cell \(m\) to the odd site of cell \(m+1\).

We Fourier transform the odd and even sites separately:
\[
\hat c_{2m-1}
=
\frac{1}{\sqrt{N_c}}
\sum_k e^{ikm}\hat c_{\mathrm{o},k},
\qquad
\hat c_{2m}
=
\frac{1}{\sqrt{N_c}}
\sum_k e^{ikm}\hat c_{\mathrm{e},k},
\]
with inverse transforms
\[
\hat c_{\mathrm{o},k}
=
\frac{1}{\sqrt{N_c}}
\sum_{m=1}^{N_c}e^{-ikm}\hat c_{2m-1},
\qquad
\hat c_{\mathrm{e},k}
=
\frac{1}{\sqrt{N_c}}
\sum_{m=1}^{N_c}e^{-ikm}\hat c_{2m}.
\]
The allowed momenta are
\[
k=\frac{2\pi n}{N_c},
\qquad
n=0,1,\ldots,N_c-1.
\]
In the thermodynamic limit, the discrete sum becomes an integral over the Brillouin zone,
\[
\frac{1}{N_c}\sum_k \longrightarrow \frac{1}{2\pi}\int_{-\pi}^{\pi}dk.
\]

We now transform the two hopping terms explicitly. For the intracell hopping,
\begin{align}
\sum_{m=1}^{N_c}\hat c_{2m-1}^\dagger\hat c_{2m}
&=
\frac{1}{N_c}
\sum_{m=1}^{N_c}
\sum_{k,k'}
e^{-ikm}e^{ik'm}
\hat c_{\mathrm{o},k}^\dagger\hat c_{\mathrm{e},k'}
\nonumber\\
&=
\frac{1}{N_c}
\sum_{k,k'}
\hat c_{\mathrm{o},k}^\dagger\hat c_{\mathrm{e},k'}
\sum_{m=1}^{N_c}e^{i(k'-k)m}
\nonumber\\
&=
\sum_k
\hat c_{\mathrm{o},k}^\dagger\hat c_{\mathrm{e},k},
\end{align}
where we used
\[
\sum_{m=1}^{N_c}e^{i(k'-k)m}=N_c\delta_{k,k'}.
\]
For the intercell hopping, note that
\[
\hat c_{2m+1}
=
\frac{1}{\sqrt{N_c}}\sum_k e^{ik(m+1)}\hat c_{\mathrm{o},k}
=
\frac{1}{\sqrt{N_c}}\sum_k e^{ikm}e^{ik}\hat c_{\mathrm{o},k}.
\]
Therefore,
\begin{align}
\sum_{m=1}^{N_c}\hat c_{2m}^\dagger\hat c_{2m+1}
&=
\frac{1}{N_c}
\sum_{m=1}^{N_c}
\sum_{k,k'}
e^{-ikm}e^{ik'm}e^{ik'}
\hat c_{\mathrm{e},k}^\dagger\hat c_{\mathrm{o},k'}
\nonumber\\
&=
\sum_k
e^{ik}
\hat c_{\mathrm{e},k}^\dagger\hat c_{\mathrm{o},k}.
\end{align}
The Hermitian conjugate gives the corresponding \(e^{-ik}\hat c_{\mathrm{o},k}^\dagger\hat c_{\mathrm{e},k}\) term.

Combining both hopping contributions, the Hamiltonian can be written as
\[
\hat H_{\mathrm{MF}}^{\mathrm{PBC}}
=
\sum_k
\Psi_k^\dagger \mathcal H(k)\Psi_k,
\qquad
\Psi_k=
\begin{pmatrix}
\hat c_{\mathrm{o},k}\\
\hat c_{\mathrm{e},k}
\end{pmatrix},
\]
where
\begin{equation}
\mathcal H(k)
=
\begin{pmatrix}
0 & q(k)\\
q^*(k) & 0
\end{pmatrix},
\qquad
q(k)=t_1^{\mathrm{eff}}+t_2^{\mathrm{eff}}e^{-ik}.
\label{eqC:Bloch}
\end{equation}
If the uniform chemical-potential shift is retained, then
\[
\mathcal H(k)
=
\mu_{\mathrm{eff}}\sigma_0
+
\begin{pmatrix}
0 & q(k)\\
q^*(k) & 0
\end{pmatrix}.
\]
Since \(\mu_{\mathrm{eff}}\sigma_0\) is proportional to the identity, it shifts both bands equally and does not affect the eigenvectors or winding number.

\subsection{Bulk spectrum and gap closing}

The eigenvalues of Eq.~\eqref{eqC:Bloch} follow from
\[
\det[\mathcal H(k)-E\mathbb I]
=
\det
\begin{pmatrix}
-E & q(k)\\
q^*(k) & -E
\end{pmatrix}
=
E^2-|q(k)|^2.
\]
The characteristic equation is
\[
E^2=|q(k)|^2,
\]
so
\[
E_\pm(k)=\pm |q(k)|.
\]
Now
\[
q(k)=t_1^{\mathrm{eff}}+t_2^{\mathrm{eff}}(\cos k-i\sin k),
\]
and hence
\begin{align}
|q(k)|^2
&=
\left[t_1^{\mathrm{eff}}+t_2^{\mathrm{eff}}\cos k\right]^2
+
\left[t_2^{\mathrm{eff}}\sin k\right]^2
\nonumber\\
&=
(t_1^{\mathrm{eff}})^2
+
2t_1^{\mathrm{eff}}t_2^{\mathrm{eff}}\cos k
+
(t_2^{\mathrm{eff}})^2(\cos^2 k+\sin^2 k)
\nonumber\\
&=
(t_1^{\mathrm{eff}})^2
+
(t_2^{\mathrm{eff}})^2
+
2t_1^{\mathrm{eff}}t_2^{\mathrm{eff}}\cos k.
\end{align}
Therefore,
\begin{equation}
E_\pm(k)
=
\pm
\sqrt{
(t_1^{\mathrm{eff}})^2
+
(t_2^{\mathrm{eff}})^2
+
2t_1^{\mathrm{eff}}t_2^{\mathrm{eff}}\cos k
}.
\label{eqC:bands}
\end{equation}
With the uniform shift restored, the bands are
\[
E_\pm(k)=2h\pm |q(k)|.
\]

The direct band separation is
\[
E_+(k)-E_-(k)=2|q(k)|.
\]
Thus the bulk gap is
\[
\Delta_{\mathrm{bulk}}=2\min_k |q(k)|.
\]
For positive effective hoppings, \(|q(k)|\) is minimized when \(\cos k=-1\), i.e. at \(k=\pi\). Hence
\begin{equation}
\Delta_{\mathrm{bulk}}
=
2|q(\pi)|
=
2|t_1^{\mathrm{eff}}-t_2^{\mathrm{eff}}|
=
2|t_2^{\mathrm{eff}}-t_1^{\mathrm{eff}}|.
\label{eqC:bulk_gap}
\end{equation}
The bulk gap closes when
\[
q(\pi)=t_1^{\mathrm{eff}}-t_2^{\mathrm{eff}}=0,
\]
or equivalently
\[
t_1^{\mathrm{eff}}=t_2^{\mathrm{eff}}.
\]
This is the SSH topological transition point.

\subsection{Winding number}

The topological invariant is determined by the phase of the complex function \(q(k)\). Writing
\[
q(k)=u(k)+iv(k),
\]
we have
\[
u(k)=\mathrm{Re}\,q(k)=t_1^{\mathrm{eff}}+t_2^{\mathrm{eff}}\cos k,
\qquad
v(k)=\mathrm{Im}\,q(k)=-t_2^{\mathrm{eff}}\sin k.
\]
As \(k\) is swept from \(-\pi\) to \(\pi\), the point \((u(k),v(k))\) traces a circle in the complex plane. The circle has radius \(|t_2^{\mathrm{eff}}|\) and is centered at \((t_1^{\mathrm{eff}},0)\).

The winding number counts how many times this circle winds around the origin:
\begin{equation}
\nu
=
\frac{1}{2\pi}
\int_{-\pi}^{\pi}
dk\,
\frac{d}{dk}\arg q(k).
\label{eqC:winding}
\end{equation}
Equivalently, using
\[
\frac{d}{dk}\arg q(k)
=
\frac{u(k)v'(k)-v(k)u'(k)}{u(k)^2+v(k)^2},
\]
one may write
\[
\nu
=
\frac{1}{2\pi}
\int_{-\pi}^{\pi}
dk\,
\frac{u(k)v'(k)-v(k)u'(k)}{|q(k)|^2}.
\]
Here
\[
u'(k)=-t_2^{\mathrm{eff}}\sin k,
\qquad
v'(k)=-t_2^{\mathrm{eff}}\cos k.
\]
Therefore,
\begin{align}
u(k)v'(k)-v(k)u'(k)
&=
(t_1^{\mathrm{eff}}+t_2^{\mathrm{eff}}\cos k)(-t_2^{\mathrm{eff}}\cos k)
-
(-t_2^{\mathrm{eff}}\sin k)(-t_2^{\mathrm{eff}}\sin k)
\nonumber\\
&=
-t_1^{\mathrm{eff}}t_2^{\mathrm{eff}}\cos k
-
(t_2^{\mathrm{eff}})^2(\cos^2 k+\sin^2 k)
\nonumber\\
&=
-t_1^{\mathrm{eff}}t_2^{\mathrm{eff}}\cos k
-
(t_2^{\mathrm{eff}})^2.
\end{align}
The sign of \(\nu\) depends on the convention \(e^{-ik}\) used in \(q(k)\). In this convention, the winding orientation is clockwise for positive \(t_2^{\mathrm{eff}}\), so the integer may be negative. The physically relevant quantity for the present discussion is the magnitude \(|\nu|\).

Geometrically, the origin lies inside the circle if and only if the radius exceeds the distance of the center from the origin:
\[
|t_2^{\mathrm{eff}}|>|t_1^{\mathrm{eff}}|.
\]
Therefore,
\begin{equation}
|\nu|
=
\begin{cases}
1, & |t_2^{\mathrm{eff}}|>|t_1^{\mathrm{eff}}|,\\[3pt]
0, & |t_2^{\mathrm{eff}}|<|t_1^{\mathrm{eff}}|.
\end{cases}
\label{eqC:winding_values}
\end{equation}
At the transition point \(t_1^{\mathrm{eff}}=t_2^{\mathrm{eff}}\), one has \(q(\pi)=0\), so the phase \(\arg q(k)\) is singular and the winding number is not defined.

\subsection{Periodic-boundary self-consistency}

The same periodic-boundary formulation gives a compact self-consistency scheme for the bulk bond-order parameters. This is useful for phase diagrams because boundary effects are absent and the finite sums are replaced by smooth momentum-space integrals.

At half filling, the lower band of Eq.~\eqref{eqC:Bloch} is occupied. Write
\[
q(k)=|q(k)|e^{-i\phi_k}.
\]
Then
\[
q^*(k)=|q(k)|e^{i\phi_k}.
\]
The Bloch Hamiltonian may be written as
\[
\mathcal H(k)
=
|q(k)|
\begin{pmatrix}
0 & e^{-i\phi_k}\\
e^{i\phi_k} & 0
\end{pmatrix}.
\]
A normalized lower-band eigenvector satisfying
\[
\mathcal H(k)|u_-(k)\rangle=-|q(k)|\,|u_-(k)\rangle
\]
is
\[
|u_-(k)\rangle
=
\frac{1}{\sqrt2}
\begin{pmatrix}
-e^{-i\phi_k}\\
1
\end{pmatrix}.
\]
Indeed,
\[
\mathcal H(k)|u_-(k)\rangle
=
\frac{|q(k)|}{\sqrt2}
\begin{pmatrix}
e^{-i\phi_k}\\
-1
\end{pmatrix}
=
-|q(k)|
\frac{1}{\sqrt2}
\begin{pmatrix}
-e^{-i\phi_k}\\
1
\end{pmatrix}.
\]

The odd-bond order parameter is the occupied-band expectation value of \(\hat c_{\mathrm{o},k}^\dagger\hat c_{\mathrm{e},k}\), averaged over \(k\). If the lower-band spinor components are denoted by
\[
u_{\mathrm{o}}(k)=-\frac{e^{-i\phi_k}}{\sqrt2},
\qquad
u_{\mathrm{e}}(k)=\frac{1}{\sqrt2},
\]
then
\[
\left\langle \hat c_{\mathrm{o},k}^\dagger\hat c_{\mathrm{e},k}\right\rangle
=
u_{\mathrm{o}}^*(k)u_{\mathrm{e}}(k)
=
-\frac{1}{2}e^{i\phi_k}
=
-\frac{q^*(k)}{2|q(k)|}.
\]
Taking the real part gives
\[
\mathrm{Re}
\left\langle \hat c_{\mathrm{o},k}^\dagger\hat c_{\mathrm{e},k}\right\rangle
=
-\frac{\mathrm{Re}\,q(k)}{2|q(k)|}
=
-\frac{t_1^{\mathrm{eff}}+t_2^{\mathrm{eff}}\cos k}{2|q(k)|}.
\]
Using evenness of the integrand,
\[
\frac{1}{2\pi}\int_{-\pi}^{\pi}dk\,f(k)=\frac{1}{\pi}\int_0^\pi dk\,f(k),
\]
we obtain
\begin{equation}
\chi_1
=
-\frac{1}{\pi}
\int_0^\pi
dk\,
\frac{t_1^{\mathrm{eff}}+t_2^{\mathrm{eff}}\cos k}{2|q(k)|}.
\label{eqC:chi1_pbc}
\end{equation}

The even-bond order parameter corresponds to the bond between the even site in cell \(m\) and the odd site in cell \(m+1\). Because of the shift by one unit cell, its momentum-space expectation value contains the phase factor \(e^{ik}\):
\[
\left\langle \hat c_{2m}^\dagger\hat c_{2m+1}\right\rangle
=
\frac{1}{N_c}\sum_k e^{ik}
\left\langle \hat c_{\mathrm{e},k}^\dagger\hat c_{\mathrm{o},k}\right\rangle.
\]
Now
\[
\left\langle \hat c_{\mathrm{e},k}^\dagger\hat c_{\mathrm{o},k}\right\rangle
=
u_{\mathrm{e}}^*(k)u_{\mathrm{o}}(k)
=
-\frac{1}{2}e^{-i\phi_k}
=
-\frac{q(k)}{2|q(k)|}.
\]
Thus
\[
e^{ik}
\left\langle \hat c_{\mathrm{e},k}^\dagger\hat c_{\mathrm{o},k}\right\rangle
=
-\frac{e^{ik}q(k)}{2|q(k)|}.
\]
Using
\[
e^{ik}q(k)
=
e^{ik}t_1^{\mathrm{eff}}+t_2^{\mathrm{eff}},
\]
its real part is
\[
\mathrm{Re}\,[e^{ik}q(k)]
=
t_1^{\mathrm{eff}}\cos k+t_2^{\mathrm{eff}}.
\]
Therefore,
\begin{equation}
\chi_2
=
-\frac{1}{\pi}
\int_0^\pi
dk\,
\frac{t_2^{\mathrm{eff}}+t_1^{\mathrm{eff}}\cos k}{2|q(k)|}.
\label{eqC:chi2_pbc}
\end{equation}

Equations~\eqref{eqC:chi1_pbc} and \eqref{eqC:chi2_pbc} must be solved together with
\[
t_1^{\mathrm{eff}}=t_1-4J_z\chi_1,
\qquad
t_2^{\mathrm{eff}}=t_2-4J_z\chi_2.
\]
The periodic-boundary self-consistency loop is therefore: choose initial values of \(\chi_1\) and \(\chi_2\); compute \(t_1^{\mathrm{eff}}\) and \(t_2^{\mathrm{eff}}\); evaluate \(q(k)\); update \(\chi_1\) and \(\chi_2\) from the integrals above; and repeat until convergence. In the thermodynamic limit, this bulk calculation agrees with the interior of the open-boundary calculation. The periodic formulation is most efficient for computing phase boundaries and winding numbers, while open boundary conditions are needed to display the edge modes directly.

\subsection{Open-boundary spectrum and edge states}

Under open boundary conditions, the closing bond is absent. In the site basis
\[
\Psi=(\hat c_1,\hat c_2,\ldots,\hat c_N)^T,
\]
the reduced mean-field Hamiltonian is represented by the tridiagonal matrix
\begin{equation}
H_{\mathrm{OBC}}
=
\begin{pmatrix}
0 & t_1^{\mathrm{eff}} & 0 & 0 & \cdots & 0\\
t_1^{\mathrm{eff}} & 0 & t_2^{\mathrm{eff}} & 0 & \cdots & 0\\
0 & t_2^{\mathrm{eff}} & 0 & t_1^{\mathrm{eff}} & \cdots & 0\\
0 & 0 & t_1^{\mathrm{eff}} & 0 & \cdots & 0\\
\vdots & \vdots & \vdots & \vdots & \ddots & t_{N-1}^{\mathrm{eff}}\\
0 & 0 & 0 & 0 & t_{N-1}^{\mathrm{eff}} & 0
\end{pmatrix},
\label{eqC:HOBC}
\end{equation}
where
\[
t_j^{\mathrm{eff}}
=
\begin{cases}
t_1^{\mathrm{eff}}, & j\ \mathrm{odd},\\[3pt]
t_2^{\mathrm{eff}}, & j\ \mathrm{even}.
\end{cases}
\]
For \(N=2N_c\), the chain is cut between unit cells. With this termination, the topological regime is
\[
|t_2^{\mathrm{eff}}|>|t_1^{\mathrm{eff}}|.
\]

The analytic edge-state wavefunctions follow from the zero-energy equation after subtracting the uniform shift:
\[
H_{\mathrm{OBC}}\psi=0.
\]
Let \(\psi_j\) be the amplitude on site \(j\). The equation on site \(j\) is
\[
t_{j-1}^{\mathrm{eff}}\psi_{j-1}+t_j^{\mathrm{eff}}\psi_{j+1}=0,
\]
with the understanding that missing boundary terms are absent. For the left edge, the first-site equation is
\[
t_1^{\mathrm{eff}}\psi_2=0.
\]
For \(t_1^{\mathrm{eff}}\neq0\), this gives
\[
\psi_2=0.
\]
The equation on site \(3\) then contains \(\psi_2\) and \(\psi_4\), and gives \(\psi_4=0\). Continuing in this way, the left zero mode has support only on odd sites:
\[
\psi_{2m}=0.
\]

The nontrivial recurrence comes from the equations on even sites. For site \(2m\),
\[
t_1^{\mathrm{eff}}\psi_{2m-1}+t_2^{\mathrm{eff}}\psi_{2m+1}=0.
\]
Thus
\[
\psi_{2m+1}
=
-\frac{t_1^{\mathrm{eff}}}{t_2^{\mathrm{eff}}}\psi_{2m-1}.
\]
Starting from the boundary amplitude \(\psi_1\), repeated iteration gives
\[
\psi_3=-\frac{t_1^{\mathrm{eff}}}{t_2^{\mathrm{eff}}}\psi_1,
\qquad
\psi_5=\left(-\frac{t_1^{\mathrm{eff}}}{t_2^{\mathrm{eff}}}\right)^2\psi_1,
\]
and in general
\begin{equation}
\psi_{2m-1}^{(L)}
=
\mathcal N_L
\left(
-\frac{t_1^{\mathrm{eff}}}{t_2^{\mathrm{eff}}}
\right)^{m-1},
\qquad
\psi_{2m}^{(L)}=0 .
\label{eqC:left_edge}
\end{equation}
The normalization constant is obtained from
\[
1
=
\sum_{m=1}^{N_c}|\psi_{2m-1}^{(L)}|^2
=
|\mathcal N_L|^2
\sum_{m=1}^{N_c}
\left|
\frac{t_1^{\mathrm{eff}}}{t_2^{\mathrm{eff}}}
\right|^{2(m-1)}.
\]
Thus
\[
|\mathcal N_L|^2
=
\frac{1-r^2}{1-r^{2N_c}},
\qquad
r=\left|\frac{t_1^{\mathrm{eff}}}{t_2^{\mathrm{eff}}}\right|,
\]
and in the thermodynamic limit,
\[
|\mathcal N_L|^2\to 1-r^2
\quad
\text{for } r<1.
\]
The state is normalizable only when
\[
\left|\frac{t_1^{\mathrm{eff}}}{t_2^{\mathrm{eff}}}\right|<1,
\]
which is exactly the topological condition.

The right-localized zero mode is obtained in the same way from the right boundary. It has support only on even sites and decays toward the left:
\[
\psi_{2m}^{(R)}
=
\mathcal N_R
\left(
-\frac{t_1^{\mathrm{eff}}}{t_2^{\mathrm{eff}}}
\right)^{N_c-m},
\qquad
\psi_{2m-1}^{(R)}=0.
\]
The decay length follows from
\[
|\psi_{2m-1}^{(L)}|
=
|\mathcal N_L|
\left|
\frac{t_1^{\mathrm{eff}}}{t_2^{\mathrm{eff}}}
\right|^{m-1}
=
|\mathcal N_L|
\exp\left[
-(m-1)\ln\left|\frac{t_2^{\mathrm{eff}}}{t_1^{\mathrm{eff}}}\right|
\right].
\]
Comparing with
\[
|\psi_{2m-1}^{(L)}|\sim e^{-(m-1)/\xi},
\]
we obtain
\begin{equation}
\xi=
\frac{1}{\ln\left|t_2^{\mathrm{eff}}/t_1^{\mathrm{eff}}\right|}.
\label{eqC:xi}
\end{equation}
The localization length diverges as \(t_1^{\mathrm{eff}}\to t_2^{\mathrm{eff}}\), exactly where the bulk gap closes.

For an infinite topological chain, the two edge states are exactly degenerate at zero energy after subtracting the uniform shift. In a finite chain, the left and right tails overlap weakly. Since each tail decays as \(|t_1^{\mathrm{eff}}/t_2^{\mathrm{eff}}|^{m}\), the splitting scales exponentially with the number of unit cells:
\begin{equation}
\delta E_{\mathrm{edge}}
\propto
\left|
\frac{t_1^{\mathrm{eff}}}{t_2^{\mathrm{eff}}}
\right|^{N_c}.
\label{eqC:edge_splitting}
\end{equation}
Restoring the uniform density shift, the edge-state energies are centered around
\[
E_{\mathrm{midgap}}=\mu_{\mathrm{eff}}=2h,
\]
and are approximately
\[
E_{\mathrm{edge}}^\pm\simeq 2h\pm \delta E_{\mathrm{edge}}.
\]
The shift changes the absolute energy but not the localization or the topological origin of the edge states.

The bulk-boundary correspondence is therefore
\begin{equation}
\begin{aligned}
|t_2^{\mathrm{eff}}|>|t_1^{\mathrm{eff}}|
&\quad\Longleftrightarrow\quad
|\nu|=1
\\
&\quad\Longleftrightarrow\quad
\text{localized open-boundary edge states}.
\end{aligned}
\label{eqC:BBC}
\end{equation}
The same effective hopping ratio controls the bulk gap, the winding number, the edge-state localization length, and the finite-size splitting.

For numerical wavefunction plots, the probability density of the \(m\)-th open-chain eigenstate is
\[
\rho_m(j)=|\psi_m(j)|^2,
\qquad
\sum_{j=1}^{N}\rho_m(j)=1.
\]
To quantify boundary localization, we define the edge weight
\[
W_{\mathrm{edge}}^{(m)}
=
\sum_{j=1}^{\ell}\rho_m(j)
+
\sum_{j=N-\ell+1}^{N}\rho_m(j),
\]
where \(\ell\) is the number of sites included near each boundary. States with large \(W_{\mathrm{edge}}^{(m)}\) and energies inside the bulk gap are identified as edge modes.

The formulas derived in this appendix are used directly in the results section. Periodic-boundary calculations use \(q(k)\), \(E_\pm(k)\), \(\Delta_{\mathrm{bulk}}\), \(\nu\), and the momentum-space self-consistency integrals. Open-boundary calculations use the tridiagonal matrix in Eq.~\eqref{eqC:HOBC}, the localization length in Eq.~\eqref{eqC:xi}, and the finite-size splitting in Eq.~\eqref{eqC:edge_splitting}.
\section{Two-Leg Ladder Geometry and Ladder Topological Spectrum}
\label{app:ladder}

This appendix extends the self-consistent SSH construction of Appendices~\ref{app:mf_decoupling} and~\ref{app:topology_edge} to a two-leg ladder. The ladder consists of two identical dimerized chains coupled by vertical rung hopping. This geometry is the minimal quasi-one-dimensional extension of the interacting SSH chain and allows us to examine how interchain hybridization modifies the bulk spectrum and the open-boundary edge-state manifold.

The central result is that, after Hartree--Fock reduction, the ladder remains a quadratic SSH-type problem. Under periodic boundary conditions, the ladder Bloch Hamiltonian separates exactly into bonding and antibonding SSH sectors. The rung hopping shifts these two sectors in energy but does not change the intrachain off-diagonal function \(q(k)\). Therefore, the topological phase boundary remains controlled by the same condition as in the single chain,
\[
|t_2^{\mathrm{eff}}|>|t_1^{\mathrm{eff}}|.
\]
The ladder doubles the number of edge states, while the rung hopping splits the edge-state doublets into bonding and antibonding combinations.

\subsection{Interacting ladder Hamiltonian}

We consider two identical dimerized chains labelled by the leg index
\[
\lambda=1,2.
\]
Each leg contains \(N\) sites. The fermionic operators are
\[
\hat c_{\lambda,j},\qquad
\hat c_{\lambda,j}^\dagger,\qquad
\hat n_{\lambda,j}=\hat c_{\lambda,j}^\dagger\hat c_{\lambda,j},
\]
where \(j=1,\ldots,N\). Along each leg, the hopping alternates according to
\[
t_j=
\begin{cases}
t_1, & j\ \mathrm{odd},\\[3pt]
t_2, & j\ \mathrm{even}.
\end{cases}
\]
Odd bonds are the intracell bonds and even bonds are the intercell bonds. In the molecular realization, the intrachain hoppings are generated by the chirality-renormalized transverse exchanges,
\[
t_{1,2}=2\widetilde J_{xy}^{(1,2)}.
\]
The rung hopping \(t_\perp\) connects site \(j\) on leg \(1\) to the same site \(j\) on leg \(2\). Since it is also dipolar in origin, it scales with the interchain separation as
\[
t_\perp\propto r_\perp^{-3}.
\]

Before mean-field decoupling, the interacting two-leg Hamiltonian may be written as
\begin{align}
\hat H_{\mathrm{ladder}}
&=
\sum_{\lambda=1}^{2}\sum_{j=1}^{N-1}
t_j
\left(
\hat c_{\lambda,j}^\dagger\hat c_{\lambda,j+1}
+
\hat c_{\lambda,j+1}^\dagger\hat c_{\lambda,j}
\right)
+
4J_z
\sum_{\lambda=1}^{2}\sum_{j=1}^{N-1}
\hat n_{\lambda,j}\hat n_{\lambda,j+1}
\nonumber\\
&\quad
+
t_\perp
\sum_{j=1}^{N}
\left(
\hat c_{1,j}^\dagger\hat c_{2,j}
+
\hat c_{2,j}^\dagger\hat c_{1,j}
\right)
+
4J_z^\perp
\sum_{j=1}^{N}
\hat n_{1,j}\hat n_{2,j}
\nonumber\\
&\quad
+
\mu_{\mathrm{bulk}}
\sum_{\lambda=1}^{2}\sum_{j=1}^{N}
\hat n_{\lambda,j}
+
E_{\mathrm{const}}^{\mathrm{ladder}} .
\label{eqD:Hladder_int}
\end{align}
The first line contains two copies of the interacting dimerized chain. The first term is the intrachain SSH hopping, while the second is the intrachain density--density interaction inherited from the Ising coupling. The second line contains the rung hopping and an optional rung density--density interaction \(J_z^\perp\). The final line contains the bulk density term and an overall constant. The constant \(E_{\mathrm{const}}^{\mathrm{ladder}}\) shifts all many-body energies uniformly and therefore does not affect the single-particle spectrum, the winding number, or the edge-state localization.

In the calculations discussed in the main text, we focus on single-particle interchain hybridization and set
\[
J_z^\perp=0.
\]
Nevertheless, keeping \(J_z^\perp\) in the derivation is useful because it shows how a rung interaction would enter the effective hopping and chemical-potential shift.

\subsection{Mean-field reduction of the ladder}

The intrachain interaction is decoupled exactly as in Appendix~\ref{app:mf_decoupling}. Since the two legs are identical, the intrachain bond-order parameters are averaged over both legs:
\[
\chi_1
=
\frac{1}{\mathcal N_1}
\sum_{\lambda=1}^{2}
\sum_{j\ \mathrm{odd}}
\left\langle
\hat c_{\lambda,j}^\dagger\hat c_{\lambda,j+1}
\right\rangle,
\]
\[
\chi_2
=
\frac{1}{\mathcal N_2}
\sum_{\lambda=1}^{2}
\sum_{j\ \mathrm{even}}
\left\langle
\hat c_{\lambda,j}^\dagger\hat c_{\lambda,j+1}
\right\rangle.
\]
Here \(\mathcal N_1\) and \(\mathcal N_2\) are the total numbers of type-\(1\) and type-\(2\) intrachain bonds over both legs.

The rung bond-order parameter is defined as
\[
\chi_\perp
=
\frac{1}{N}
\sum_{j=1}^{N}
\left\langle
\hat c_{1,j}^\dagger\hat c_{2,j}
\right\rangle .
\]
It measures the interchain coherence across the rungs. If \(J_z^\perp=0\), \(\chi_\perp\) can still be evaluated as an observable, but it does not feed back into the Hamiltonian.

The only new Hartree--Fock step relative to the single chain is the rung interaction. On a given rung,
\[
4J_z^\perp \hat n_{1,j}\hat n_{2,j}
=
4J_z^\perp
\hat c_{1,j}^\dagger\hat c_{1,j}
\hat c_{2,j}^\dagger\hat c_{2,j}.
\]
At half filling,
\[
\langle \hat n_{1,j}\rangle
=
\langle \hat n_{2,j}\rangle
=
\frac12,
\qquad
\left\langle
\hat c_{1,j}^\dagger\hat c_{2,j}
\right\rangle
=
\chi_\perp .
\]
Using the same Hartree--Fock decomposition as in Appendix~\ref{app:mf_decoupling}, we obtain
\begin{align}
4J_z^\perp\hat n_{1,j}\hat n_{2,j}
&\approx
2J_z^\perp
\left(
\hat n_{1,j}
+
\hat n_{2,j}
\right)
-
4J_z^\perp\chi_\perp
\left(
\hat c_{1,j}^\dagger\hat c_{2,j}
+
\hat c_{2,j}^\dagger\hat c_{1,j}
\right)
\nonumber\\
&\quad
-
J_z^\perp
+
4J_z^\perp\chi_\perp^2 .
\label{eqD:rung_decoupling}
\end{align}
The first term is the rung Hartree shift, the second is the rung Fock correction, and the final two terms are constants.

Combining the bare hopping terms with the Fock corrections gives
\begin{equation}
t_1^{\mathrm{eff}}
=
t_1-4J_z\chi_1,
\qquad
t_2^{\mathrm{eff}}
=
t_2-4J_z\chi_2,
\qquad
t_\perp^{\mathrm{eff}}
=
t_\perp-4J_z^\perp\chi_\perp .
\label{eqD:effective_hoppings}
\end{equation}
The density shifts combine in the same way as in the single chain. The intrachain Hartree term cancels the \(-4J_z\) part of \(\mu_{\mathrm{bulk}}\). The rung Hartree term contributes \(2J_z^\perp\) per site, because each site belongs to one rung. Therefore,
\[
\mu_{\mathrm{eff}}^{\mathrm{ladder}}
=
\mu_{\mathrm{bulk}}+4J_z+2J_z^\perp
=
2h+2J_z^\perp .
\]
For the case used in the main text,
\[
J_z^\perp=0,
\qquad
t_\perp^{\mathrm{eff}}=t_\perp,
\qquad
\mu_{\mathrm{eff}}^{\mathrm{ladder}}=2h.
\]
The uniform density shift is proportional to the identity in the single-particle Hamiltonian. It is therefore subtracted in the topological analysis and restored only when absolute energies are plotted.

After subtracting the uniform shift and dropping constants, the quadratic mean-field ladder Hamiltonian is
\begin{align}
\hat H_{\mathrm{ladder}}^{\mathrm{MF}}
&=
\sum_{\lambda=1}^{2}\sum_{j=1}^{N-1}
t_j^{\mathrm{eff}}
\left(
\hat c_{\lambda,j}^\dagger\hat c_{\lambda,j+1}
+
\hat c_{\lambda,j+1}^\dagger\hat c_{\lambda,j}
\right)
\nonumber\\
&\quad
+
t_\perp^{\mathrm{eff}}
\sum_{j=1}^{N}
\left(
\hat c_{1,j}^\dagger\hat c_{2,j}
+
\hat c_{2,j}^\dagger\hat c_{1,j}
\right),
\label{eqD:Hladder_MF}
\end{align}
where
\[
t_j^{\mathrm{eff}}
=
\begin{cases}
t_1^{\mathrm{eff}}, & j\ \mathrm{odd},\\[3pt]
t_2^{\mathrm{eff}}, & j\ \mathrm{even}.
\end{cases}
\]

\subsection{Open-boundary matrix representation}

Under open boundary conditions, no bond connects site \(N\) back to site \(1\) along either leg. We order the single-particle basis as
\[
\Psi
=
(
\hat c_{1,1},
\hat c_{1,2},
\ldots,
\hat c_{1,N},
\hat c_{2,1},
\hat c_{2,2},
\ldots,
\hat c_{2,N}
)^T .
\]
In this basis, the ladder Hamiltonian takes the block form
\begin{equation}
H_{\mathrm{ladder}}^{\mathrm{OBC}}
=
\begin{pmatrix}
H_{\mathrm{SSH}} & t_\perp^{\mathrm{eff}}\mathbb I_N\\
t_\perp^{\mathrm{eff}}\mathbb I_N & H_{\mathrm{SSH}}
\end{pmatrix}.
\label{eqD:OBC_block}
\end{equation}
Here \(H_{\mathrm{SSH}}\) is the \(N\times N\) open-boundary SSH matrix of one leg:
\[
(H_{\mathrm{SSH}})_{j,j+1}
=
(H_{\mathrm{SSH}})_{j+1,j}
=
\begin{cases}
t_1^{\mathrm{eff}}, & j\ \mathrm{odd},\\[3pt]
t_2^{\mathrm{eff}}, & j\ \mathrm{even}.
\end{cases}
\]
The off-diagonal blocks are proportional to the identity because the rung hopping connects only equal site indices on the two legs:
\[
(1,j)\leftrightarrow (2,j).
\]
This real-space block structure makes clear that the ladder is composed of two SSH chains coupled by vertical rung hybridization.

\subsection{Periodic-boundary Bloch Hamiltonian}

For the bulk ladder spectrum, we impose periodic boundary conditions along the chains. As in Appendix~\ref{app:topology_edge}, odd and even sites are Fourier transformed separately on each leg:
\[
\hat c_{\lambda,2m-1}
=
\frac{1}{\sqrt{N_c}}
\sum_k e^{ikm}\hat c_{\lambda,\mathrm{o},k},
\qquad
\hat c_{\lambda,2m}
=
\frac{1}{\sqrt{N_c}}
\sum_k e^{ikm}\hat c_{\lambda,\mathrm{e},k},
\]
where \(N_c=N/2\). The four-component Bloch spinor is
\[
\Psi_k
=
(
\hat c_{1,\mathrm{o},k},
\hat c_{1,\mathrm{e},k},
\hat c_{2,\mathrm{o},k},
\hat c_{2,\mathrm{e},k}
)^T .
\]
Using the single-chain result
\[
q(k)=t_1^{\mathrm{eff}}+t_2^{\mathrm{eff}}e^{-ik},
\]
the intrachain hopping on each leg contributes the same SSH block
\[
H_{\mathrm{SSH}}(k)
=
\begin{pmatrix}
0 & q(k)\\
q^*(k) & 0
\end{pmatrix}.
\]
The rung hopping connects equal sublattice indices on the two legs. Therefore, in momentum space,
\[
t_\perp^{\mathrm{eff}}\sum_j
(
\hat c_{1,j}^\dagger\hat c_{2,j}
+
\mathrm{h.c.}
)
=
t_\perp^{\mathrm{eff}}\sum_k
(
\hat c_{1,\mathrm{o},k}^\dagger\hat c_{2,\mathrm{o},k}
+
\hat c_{1,\mathrm{e},k}^\dagger\hat c_{2,\mathrm{e},k}
+
\mathrm{h.c.}
).
\]
There is no extra phase factor in the rung term because it connects sites within the same unit cell.

The periodic ladder Hamiltonian can therefore be written as
\[
\hat H_{\mathrm{ladder}}^{\mathrm{PBC}}
=
\sum_k
\Psi_k^\dagger
\mathcal H_{\mathrm{ladder}}(k)
\Psi_k,
\]
with
\begin{equation}
\mathcal H_{\mathrm{ladder}}(k)
=
\begin{pmatrix}
0 & q(k) & t_\perp^{\mathrm{eff}} & 0\\
q^*(k) & 0 & 0 & t_\perp^{\mathrm{eff}}\\
t_\perp^{\mathrm{eff}} & 0 & 0 & q(k)\\
0 & t_\perp^{\mathrm{eff}} & q^*(k) & 0
\end{pmatrix}.
\label{eqD:ladder_Bloch}
\end{equation}
If the uniform density shift is retained, one adds
\[
\mu_{\mathrm{eff}}^{\mathrm{ladder}}\mathbb I_{4\times4}
\]
to Eq.~\eqref{eqD:ladder_Bloch}.

\subsection{Bonding--antibonding block diagonalization}

The rung hopping acts only in the leg space and is proportional to the identity in the odd-even sublattice space. Therefore, the ladder Hamiltonian can be diagonalized in the leg sector by forming bonding and antibonding combinations:
\[
\hat c_{+,\mathrm{o},k}
=
\frac{1}{\sqrt2}
(
\hat c_{1,\mathrm{o},k}
+
\hat c_{2,\mathrm{o},k}
),
\qquad
\hat c_{-,\mathrm{o},k}
=
\frac{1}{\sqrt2}
(
\hat c_{1,\mathrm{o},k}
-
\hat c_{2,\mathrm{o},k}
),
\]
and
\[
\hat c_{+,\mathrm{e},k}
=
\frac{1}{\sqrt2}
(
\hat c_{1,\mathrm{e},k}
+
\hat c_{2,\mathrm{e},k}
),
\qquad
\hat c_{-,\mathrm{e},k}
=
\frac{1}{\sqrt2}
(
\hat c_{1,\mathrm{e},k}
-
\hat c_{2,\mathrm{e},k}
).
\]
Equivalently, the unitary transformation in leg space is
\[
U_{\mathrm{leg}}
=
\frac{1}{\sqrt2}
\begin{pmatrix}
1 & 1\\
1 & -1
\end{pmatrix}.
\]
This diagonalizes the rung matrix
\[
\begin{pmatrix}
0 & t_\perp^{\mathrm{eff}}\\
t_\perp^{\mathrm{eff}} & 0
\end{pmatrix}
\longrightarrow
\begin{pmatrix}
t_\perp^{\mathrm{eff}} & 0\\
0 & -t_\perp^{\mathrm{eff}}
\end{pmatrix}.
\]
Since the intrachain SSH block is identical on both legs, it remains unchanged by this transformation. Hence,
\[
\mathcal H_{\mathrm{ladder}}(k)
\longrightarrow
\mathcal H_+(k)\oplus\mathcal H_-(k),
\]
where
\begin{equation}
\mathcal H_\eta(k)
=
\begin{pmatrix}
\eta t_\perp^{\mathrm{eff}} & q(k)\\
q^*(k) & \eta t_\perp^{\mathrm{eff}}
\end{pmatrix},
\qquad
\eta=\pm1 .
\label{eqD:Heta}
\end{equation}
Thus the ladder is exactly equivalent, at the quadratic level, to two SSH Hamiltonians shifted in energy by \(\pm t_\perp^{\mathrm{eff}}\). The rung hopping does not modify \(q(k)\).

The eigenvalues of each block satisfy
\[
\det[\mathcal H_\eta(k)-E\mathbb I]
=
(E-\eta t_\perp^{\mathrm{eff}})^2-|q(k)|^2=0.
\]
Therefore,
\begin{equation}
E_{\eta,\pm}(k)
=
\eta t_\perp^{\mathrm{eff}}
\pm
|q(k)|,
\qquad
\eta=\pm1 .
\label{eqD:ladder_bands}
\end{equation}
Restoring the uniform shift gives
\[
E_{\eta,\pm}(k)
=
\mu_{\mathrm{eff}}^{\mathrm{ladder}}
+
\eta t_\perp^{\mathrm{eff}}
\pm
|q(k)|.
\]
The intrinsic SSH gap inside each bonding or antibonding sector is
\[
\Delta_{\mathrm{SSH}}
=
2|t_2^{\mathrm{eff}}-t_1^{\mathrm{eff}}|.
\]
The rung hopping shifts the two SSH copies relative to one another but does not change this gap.

\subsection{Ladder winding number}

The winding number of each bonding sector is determined by the same function \(q(k)\) as in the single chain:
\[
\nu_\eta
=
\frac{1}{2\pi}
\int_{-\pi}^{\pi}dk\,
\frac{d}{dk}\arg q(k),
\qquad
\eta=\pm1.
\]
Because both sectors contain the same \(q(k)\),
\[
\nu_+=\nu_-=\nu_{\mathrm{SSH}}.
\]
The total ladder winding number is therefore
\begin{equation}
\nu_{\mathrm{ladder}}
=
\nu_++\nu_-
=
2\nu_{\mathrm{SSH}}.
\label{eqD:nu_ladder}
\end{equation}
In terms of the hopping amplitudes,
\[
|\nu_{\mathrm{ladder}}|
=
\begin{cases}
2, & |t_2^{\mathrm{eff}}|>|t_1^{\mathrm{eff}}|,\\[3pt]
0, & |t_2^{\mathrm{eff}}|<|t_1^{\mathrm{eff}}|.
\end{cases}
\]
The sign depends on the Fourier convention. The magnitude is the physically relevant quantity for distinguishing the trivial and topological ladder phases.

\subsection{Edge-state splitting and protection condition}

In the decoupled limit \(t_\perp^{\mathrm{eff}}=0\), the ladder is simply two independent SSH chains. If each leg is in the topological phase, each chain contributes one left edge state and one right edge state. The ladder therefore contains four edge states in total.

When \(t_\perp^{\mathrm{eff}}\neq0\), the two edge states located at the same boundary hybridize through the rung hopping. Let \(\ket{L,1}\) and \(\ket{L,2}\) denote the left edge states on legs \(1\) and \(2\). Projecting the rung hopping onto this two-dimensional edge subspace gives
\[
H_{\mathrm{edge}}^{(L)}
=
\begin{pmatrix}
0 & t_\perp^{\mathrm{eff}}\\
t_\perp^{\mathrm{eff}} & 0
\end{pmatrix}.
\]
The eigenstates are
\[
\ket{L,+}
=
\frac{1}{\sqrt2}
(\ket{L,1}+\ket{L,2}),
\qquad
\ket{L,-}
=
\frac{1}{\sqrt2}
(\ket{L,1}-\ket{L,2}),
\]
with energies
\begin{equation}
E_{\mathrm{edge}}^\pm
\simeq
\pm t_\perp^{\mathrm{eff}}.
\label{eqD:edge_split}
\end{equation}
The same splitting occurs at the right boundary. Restoring the uniform density shift gives
\[
E_{\mathrm{edge}}^\pm
\simeq
\mu_{\mathrm{eff}}^{\mathrm{ladder}}
\pm
t_\perp^{\mathrm{eff}}.
\]

For a finite ladder, there is also an exponentially small left-right overlap along each leg. As in Appendix~\ref{app:topology_edge}, this contribution scales as
\[
\delta E_{\mathrm{leg}}
\propto
\left|
\frac{t_1^{\mathrm{eff}}}{t_2^{\mathrm{eff}}}
\right|^{N_c}.
\]
The rung splitting is the dominant edge splitting when
\[
|t_\perp^{\mathrm{eff}}|
\gg
\left|
\frac{t_1^{\mathrm{eff}}}{t_2^{\mathrm{eff}}}
\right|^{N_c}.
\]

The rung-split edge modes remain spectrally isolated only if they do not merge with the bulk continuum. In each SSH sector, the distance from midgap to the nearest bulk band edge is
\[
|t_2^{\mathrm{eff}}-t_1^{\mathrm{eff}}|.
\]
Therefore, a necessary in-gap condition is
\begin{equation}
|t_\perp^{\mathrm{eff}}|
<
|t_2^{\mathrm{eff}}-t_1^{\mathrm{eff}}|.
\label{eqD:edge_condition}
\end{equation}
For clearly isolated edge modes in numerical spectra, we use the stronger practical criterion
\[
|t_\perp^{\mathrm{eff}}|
\lesssim
\frac12
|t_2^{\mathrm{eff}}-t_1^{\mathrm{eff}}|.
\]
This keeps the rung-split edge levels well separated from the nearest bulk band. In the molecular platform, the criterion can be satisfied by increasing \(r_\perp\), since \(t_\perp\propto r_\perp^{-3}\).

\subsection{Ladder self-consistency}

The open-boundary self-consistency procedure is a direct extension of the single-chain calculation. Diagonalizing Eq.~\eqref{eqD:OBC_block}, let
\[
\psi_m(\lambda,j)
\]
be the amplitude of the \(m\)-th eigenstate on leg \(\lambda\) and site \(j\). The ladder has \(2N\) sites, so at half filling the lowest \(N\) single-particle states are occupied. The intrachain bond orders are
\[
\chi_1
=
\frac{1}{\mathcal N_1}
\sum_{m\in\mathrm{occ}}
\sum_{\lambda=1}^{2}
\sum_{j\ \mathrm{odd}}
\psi_m^*(\lambda,j)\psi_m(\lambda,j+1),
\]
\[
\chi_2
=
\frac{1}{\mathcal N_2}
\sum_{m\in\mathrm{occ}}
\sum_{\lambda=1}^{2}
\sum_{j\ \mathrm{even}}
\psi_m^*(\lambda,j)\psi_m(\lambda,j+1).
\]
If \(J_z^\perp\neq0\), the rung bond order must also be updated:
\[
\chi_\perp
=
\frac{1}{N}
\sum_{m\in\mathrm{occ}}
\sum_{j=1}^{N}
\psi_m^*(1,j)\psi_m(2,j).
\]
These quantities are substituted back into
\[
t_1^{\mathrm{eff}}=t_1-4J_z\chi_1,
\qquad
t_2^{\mathrm{eff}}=t_2-4J_z\chi_2,
\qquad
t_\perp^{\mathrm{eff}}=t_\perp-4J_z^\perp\chi_\perp,
\]
and the procedure is iterated until convergence. If \(J_z^\perp=0\), then
\[
t_\perp^{\mathrm{eff}}=t_\perp,
\]
so only the intrachain bond orders \(\chi_1\) and \(\chi_2\) enter the self-consistency loop.

The relation between the single-chain and two-leg ladder descriptions is summarized below:
\begin{ruledtabular}
\begin{tabular}{lll}
\textbf{Quantity} & \textbf{Single chain} & \textbf{Two-leg ladder} \\
\hline
Degrees of freedom
& $N$ sites
& $2N$ sites \\
Mean-field hoppings
& $t_1^{\mathrm{eff}},\,t_2^{\mathrm{eff}}$
& $t_1^{\mathrm{eff}},\,t_2^{\mathrm{eff}},\,t_\perp^{\mathrm{eff}}$ \\
Interaction channels
& Intrachain $J_z$
& Intrachain $J_z$ and rung $J_z^\perp$ \\
PBC Bloch Hamiltonian
& $2\times2$ SSH matrix
& $4\times4$ ladder matrix \\
Block structure
& Single SSH block
& Bonding/antibonding SSH blocks \\
Bulk bands
& $E_\pm=\mu_{\mathrm{eff}}\pm |q(k)|$
& $E_{\eta,\pm}=\mu_{\mathrm{eff}}^{\mathrm{ladder}}+\eta t_\perp^{\mathrm{eff}}\pm |q(k)|$ \\
Topological invariant
& $|\nu|=1$
& $|\nu_{\mathrm{ladder}}|=2$ \\
OBC edge states
& Two edge states
& Four edge states \\
Effect of rung hopping
& Not applicable
& Splits edge states by $\approx 2t_\perp^{\mathrm{eff}}$ \\
Topological condition
& $|t_2^{\mathrm{eff}}|>|t_1^{\mathrm{eff}}|$
& Same intrachain condition \\
\end{tabular}
\end{ruledtabular}
\vspace{0.4em}
{\small \textbf{Table~D.1:} Comparison between the single-chain SSH 
mean-field model and its two-leg ladder extension.}

Thus the ladder preserves the SSH topology of the individual chains while enriching the edge spectrum through interchain hybridization. The rung hopping does not shift the topological phase boundary because it does not modify \(q(k)\). Instead, it splits the four open-boundary edge states into bonding and antibonding pairs. In the weak-rung regime, these modes remain localized and spectrally isolated, giving a controlled ladder extension of the interacting chiral SSH chain.

\end{document}